\documentclass[pra,twocolumn,groupedaddress]{revtex4-1}

\usepackage{graphicx}% Include figure files
\usepackage{dcolumn}% Align table columns on decimal point
\usepackage{bm}% bold math
\usepackage{epsfig}
\usepackage{amsmath}
\usepackage{amssymb}
\usepackage{color}
\usepackage{bbm}
\usepackage{braket}
\usepackage{float}
\usepackage{enumitem}
\usepackage{tensor}
\usepackage{xcolor}

\newcommand{\iac}[1]{{\color{black}#1}}
\newcommand{\ddb}[1]{{\color{black}#1}}
\usepackage[colorlinks=true,citecolor=blue,linkcolor=blue,urlcolor=blue]{hyperref}

\begin{document}

%\title{Meta-Rabi coupling between ultra-strong coupling polaritons and intersubband transitions}
%\title{Polariton-mediated cavity QED in the ultra-strong coupling regime}
%\title{Ultrastrong coupling tomography from polariton-matter interactions with intersubband quantum wells\\{\color{red} Full tomography of polaritonic states in the ultrastrong coupling regime: the case of intersubband quantum wells.}}
%\title{Modification of light emission processes in the vacuum of ultrastrongly coupled systems}
\title{Light-matter interactions in the vacuum of ultra-strongly coupled systems}

\author{Daniele De Bernardis$^{1,2,3}$, Gian Marcello Andolina$^2$, and Iacopo Carusotto$^1$}
\affiliation{$^1$Pitaevskii BEC Center, INO-CNR and Dipartimento di Fisica, Universit\`a di Trento
I-38123 Trento, Italy}
\affiliation{$^2$JEIP, UAR 3573 CNRS, Collège de France, PSL Research University, 11 Place Marcelin Berthelot,  F-75321 Paris, France}
\affiliation{$^3$National Institute of Optics [Consiglio Nazionale delle Ricerche CNR–INO], care of European Laboratory for Non-Linear Spectroscopy (LENS), Via Nello Carrara 1, Sesto Fiorentino, 50019, Italy}

\date{\today}

%%%%
\begin{abstract} 
%We prose a method to measure the light-matter hybridization in the ultrastrong coupling regime of cavity QED by measuring the Rabi splitting of a auxiliary emitter coupled to the cavity polaritons. The specific system under consideration consists of two quantum wells placed within a planar cavity. Here, their intersubband transitions are coupled with the transverse magnetic modes of the cavity and by controlling the doping amount we can assume that one is ultrastrongly coupled to the cavity while the other is only strongly coupled. We show that the Rabi splitting of the weakly doped quantum well is directly proportional to the Hopfield coefficients of the ultrastrongly coupled quantum well. By considering only standard transmission/reflection spectroscopy, we propose a protocol to obtain a full tomography of the polaritonic vacuum which is realizable in current intersubband transitions experiments.
%Ultrastrong coupling between light and matter can reveal inherent quantum mechanical phenomena, such as quantum vacuum fluctuations. 
We theoretically study how the peculiar properties of the vacuum state of an ultra-strongly coupled system can affect basic light-matter interaction processes. In this unconventional electromagnetic environment, an additional emitter no longer couples to the bare cavity photons, but rather to the polariton modes emerging from the ultra-strong coupling. \iac{As such,} the effective light-matter interaction strength is sensitive to the properties of the distorted vacuum state. Different interpretations of our predictions in terms of modified quantum fluctuations in the vacuum state and of radiative reaction in classical electromagnetism are critically discussed.
Whereas our discussion is focused on the experimentally most relevant case of intersubband polaritons in semiconductor devices, our framework is fully general and applies to generic material systems.
% modifying its interaction strength.
% The new Rabi splitting emerging from this resonant interaction is now sensitive to the dressed vacuum allowing for a direct measurement of the hybridization Hopfield coefficients.
% We specify our discussion to the case of intersubband polaritons even if the formalism is completely general.
% %We then propose to use it as a direct measurement of the hybridization Hopfield coefficients that are here represented in the form of a single hybridization angle.
% %Our model is constituted by two quantum wells embedded in a planar cavity, whose intersubband transitions are coupled with the cavity modes. By tuning the electronic doping of the quantum wells, we set in the case where one well is ultrastrongly coupled while the other is assumed only strongly coupled, taking the role of the emitter. 
% Finally, by analyzing this problem through the classical Maxwell equations in dielectric media we show that our findings can be re-interpreted equivalently from a classical electrodynamics perspective.
% This analysis reveals an unexpected role of the electrostatic forces that is crucial in the modification of the emitter's properties. 
% In these regards, the effect of the non-trivial vacuum can be mostly re-interpreted by purely electrostatic means.
% %{\color{red} The real role of quantum vacuum becomes mostly a matter of perspective and representation}.
\end{abstract}
 
\maketitle

\section{Introduction}
The non-empty nature of the quantum vacuum is among the most fascinating effects emerging from quantum mechanics and quantum field theories~\cite{milonni1994quantum}.
Fundamental effects of atomic physics such as the Lamb shift~\cite{Bethe_PhysRev.72.339}, the spontaneous emission~\cite{ milonni_why_atom_radiate_10.1119/1.13886}, and the vacuum-field Rabi oscillations~\cite{burstein2012confined,Mondragon_PhysRevLett.51.550,agarwal_PhysRevLett.53.1732, Berman_book}
%, Thompson_PhysRevLett.68.1132
can be traced back to the quantum fluctuations of the electromagnetic field in the vacuum state. 
However, the quantum vacuum reveals itself also at a more macroscopic scale, for instance through the Casimir forces \cite{Kimball_A_Milton_2004, Klimchitskaya_RevModPhys.81.1827, Wang2021_nature_casimir_geometry}, bringing the idea that this intriguing feature of quantum physics can be exploited for nano-manipulation and nano-mechanical devices~\cite{S_Haroche_1991, capasso_4159963,GongCorradoMahbubSheldenMunday+2021+523+536}.
%At a macroscopic level, an intense theoretical and experimental activity has been devoted to Casimir forces between metallic or dielectric objects in different geometries~\cite{Kimball_A_Milton_2004, Klimchitskaya_RevModPhys.81.1827, Wang2021_nature_casimir_geometry}, with exciting perspectives for nano-manipulation and nano-mechanical devices~\cite{S_Haroche_1991, capasso_4159963,GongCorradoMahbubSheldenMunday+2021+523+536}. More recently, a strong interest is being paid also to the emission of quantum radiation by moving neutral objects via the dynamical Casimir effect~\cite{Lambrecht:JPhys2012,birrell1984quantum} and to the corresponding radiation friction forces~\cite{Kardar:RMP1999}, effects which are still awaiting a complete experimental verification. 

In the last years, the physics of the quantum vacuum \iac{has started attracting} a growing interest also from the point of view of condensed matter physics \iac{as an} innovative way to manipulate the \iac{microscopic} interaction mechanisms \iac{between electrons} and\iac{, thus,} induce new states of quantum matter~\cite{ebbesen_ciuti_review, Jacqueline_mohammad_review}.
A crucial ingredient here is the capability to reach the \emph{ultrastrong coupling} (USC) regime of light-matter interactions~\cite{Ciuti_PhysRevB.72.115303,nori_USC_review, Solano_USC_RevModPhys.91.025005} where the extremely large value of the coupling strength of polarizable emitters to the electromagnetic field leads to a significant distortion of the properties of the electromagnetic vacuum and\iac{, in particular, of} its quantum fluctuations.

In this work, we give a new twist to this research by investigating how the peculiar properties of the USC vacuum affect basic light-matter interaction processes involving another emitter used as a probe. Experimentally observable consequences of this physics are highlighted, such as marked modifications of the vacuum-field Rabi oscillations and of the spontaneous emission rate. While these features are naturally understood as an experimentally accessible evidence of the distorted %presence of virtual excitations in the 
quantum vacuum state, alternative interpretations %explanations 
based on classical electromagnetism, fluctuation-dissipation theorems, and electrostatic forces are also proposed and critically discussed.

Modulo straightforward modifications, our framework applies to generic systems where the USC can be achieved~\cite{nori_USC_review, Solano_USC_RevModPhys.91.025005}, from cyclotron excitations in metallic resonators~\cite{scalari_science_2012,scalari_LandauPol_USC_10.1063/1.4795543,Faist_magneto_transport_2019NatPhys}, to Josephson-junction-based devices coupled to superconducting microwave resonators~\cite{yoshihara_circuitQED_beyond_USC2017, tomonaga2023_2qubit_USC}, \iac{and} excitons in 2D materials or \iac{heterostructures}~\cite{Rose_USC_2D_mat_advoptmat_2022, Papadimopoulos:17, theodoros_PhysRevApplied.19.064072, anantharaman2023ultrastrong}. For the sake of concreteness, in this work, we focus \iac{however} our attention on a \iac{specific} semiconductor-based platform based on intersubband (ISB) transitions in quantum wells (QW), where USC was first predicted~\cite{Ciuti_PhysRevB.72.115303} and observed~\cite{tredicucci_PhysRevB.79.201303, Sirtori_USC_MIR_doi:10.1063/1.3598432, Todorov_2010_USC_polariton_Dot_PhysRevLett.105.196402, Dietze_USC_doi:10.1063/1.4830092, Askenazi_sirtori_2014_USC_Reststrahlen_band,cortese2021excitons, pisani_USCtransport_ISB_natcom2023}. Such a platform still remains among the most promising platforms for the study of quantum vacuum effects~\cite{ Deliberato_PhysRevLett.98.103602, Ciuti_subcycle_2021_PhysRevLett.126.177404, ciuti_subcycle_natphot_2021} \iac{and appears as a most promising choice for the experimental verification of our predictions.}

% \iac{The specific geometry under investigation is sketched in Fig.~\ref{fig:1}(a): it is} based on a planar metallic cavity mode ultrastrongly coupled to an ISB transition in a heavily doped QW, in the following called the \emph{dresser}.
% %In specific, the USC regime for intersubband polaritons is obtained by inserting in the cavity a heavily doped quantum well (QW), in the following called the \emph{dresser}. 
% The dressed vacuum of the USC regime is then used to non-perturbatively influence the coupling to the electromagnetic field of another QW, called the \emph{emitter}. This latter QW is taken to be much less doped, so that the coupling to light of its own ISB transition  is far below the ultra-strong coupling regime.
% Depending on its strong vs. weak coupling to light,  
% %, but still well in the strong coupling one. In this case, the basic light-matter interaction process consists of a vacuum-field Rabi oscillation between an electronic excitation in the emitter QW and the cavity-dresser polariton modes resulting from the USC of the dresser QW and the cavity mode, and has  a straightforward spectroscopic signature as an additional anticrossing in the cavity transmission spectrum. In particular, 
% the distortion of the vacuum state in the USC regime is visible as a strong reinforcement (suppression) of the vacuum-field Rabi oscillation frequency or of the spontaneous emission rate when the emitter is resonant with lower (upper) polariton branch.

The article is organized as follows:
in Sec. \ref{sec:general_framework} we introduce \iac{our general framework based on polaritonic correlation function \ddb{and master equation} to describe the modified electromagnetic environment in the ultra-strong light matter coupling regime and its coupling to a weak emitter.}
%dressing of the electromagnetic environment with matter in ultrastrong coupling and probing it with an emitter only weakly coupled. 
In Sec. \ref{sec:cavity-dresser_ISB_pol} we apply our general scheme to describe the specific setup of intersubband polaritons \iac{and we formulate the system dynamics in terms of an emitter coupled to a polaritonic environment.
In Sec. \ref{sec:quantum_fluct_USC_vacuum} we show how \ddb{the dipole-dipole} interactions and the spontaneous emission properties of the emitter are related to the \iac{quantum fluctuations in the non-trivial} USC quantum vacuum.
In Sec. \ref{sec:relation_classical_dielectric} we reinterpret \iac{the results of} the previous sections in terms of classical dielectric theory, explicitly showing the equivalence between the quantum and the classical Maxwell equation frameworks.
Finally in Sec. \ref{sec:conclusion} we draw our conclusions.
}

%\section{Model}
%We consider the paradigmatic cavity QED setup represented in Fig. \ref{fig:1}, 

\section{General theory}
\label{sec:general_framework}
\subsection{Quantum electrodynamics with polarizable media}

We consider a system composed by two polarizable dielectrics, called dresser and emitter, embedded \iac{in} the quantized electromagnetic environment of a resonant cavity. In the dipole gauge picture, dubbed $\textbf{d}\cdot \textbf{E}$ in the literature~\cite{Cohen-Tannoudji:113864,craig1998molecular}, \iac{the Hamiltonian can be generically expressed in the form}
\begin{equation}\label{eq:Ham_general}
\begin{split}
&H = H_{\rm d} + H_e +
\\
& \int d^3 r \left[ \frac{\left( \textbf{D}(\textbf{r}) - \textbf{P}_{\rm d}(\textbf{r}) - \textbf{P}_{e}(\textbf{r}) \right)^2}{2 \epsilon_0} - \frac{\epsilon_0 c^2}{2}\textbf{A}(\textbf{r})\cdot \nabla^2 \textbf{A}(\textbf{r})  \right]
\end{split}
\end{equation}
\iac{where $H_{{\rm d},e}$} are the dresser and emitter Hamiltonians. Both the electric displacement field $\textbf{D}(\textbf{r})$ and the vector potential $\textbf{A}(\textbf{r})$ satisfy the transversality condition $\vec{\nabla}\cdot \textbf{D}(\textbf{r}) =\vec{\nabla}\cdot\textbf{A}(\textbf{r})= 0$; and $\textbf{P}_{\rm d,e}(\textbf{r})$ % \textbf{P}_{e}(\textbf{r})$ 
are the dresser and emitter polarization densities. $\epsilon_0$ is the vacuum dielectric permittivity and $c$ is the speed of light.

In this work we always assume that the dresser and emitter are spatially separated, so their polarization densities have zero overlap, $\int d^3 r\,\textbf{P}_{\rm d}(\textbf{r}) \cdot  \textbf{P}_{e}(\textbf{r}) = 0$.
The general Hamiltonian can then be rewritten as
\begin{equation}\label{eq:Ham_general_expanded}
\begin{split}
H = & H_e  - \int d^3 r\,  \frac{\textbf{D}(\textbf{r})\cdot \textbf{P}_{e}(\textbf{r})}{ \epsilon_0} + \int d^3 r  \frac{ \textbf{P}^2_{e}(\textbf{r})}{2 \epsilon_0} + H_{\rm pol},
\end{split}
\end{equation}
where we define the polariton Hamiltonian \iac{emerging from the USC of the dresser to the cavity mode} as
\begin{equation}\label{eq:Ham_general_expanded}
\begin{split}
&H_{\rm pol} = H_{\rm d} + \int d^3 r\,\left[ \frac{ \textbf{D}^2(\textbf{r})}{2 \epsilon_0} - \frac{\epsilon_0 c^2}{2}\textbf{A}(\textbf{r})\cdot \nabla^2 \textbf{A}(\textbf{r})  \right]
\\ 
& - \int d^3 r  \frac{\textbf{D}(\textbf{r})\cdot \textbf{P}_{\rm d}(\textbf{r})}{ \epsilon_0} + \int d^3 r  \frac{ \textbf{P}^2_{\rm d}(\textbf{r})}{2 \epsilon_0}.
\end{split}
\end{equation}

As it is clearly visible from the interaction part of the Hamiltonian
\begin{equation}\label{eq:ham_emitter_D_general}
H_{e/{\rm d},\,I} = - \int d^3 r~  \frac{\textbf{D}(\textbf{r})\cdot \textbf{P}_{e/{\rm d}}(\textbf{r})}{\epsilon_0} + \int d^3 r ~ \frac{\textbf{P}^2_{e/{\rm d}}(\textbf{r})}{2 \epsilon_0}
\end{equation}
in the dipole gauge representation, \iac{both} the emitter and \iac{the} dresser \iac{only} interact with the electric displacement field, without any direct Coulomb coupling \iac{between them} \cite{Cohen-Tannoudji:113864}. Notice that the $P^2$-term must be \iac{also} included as a part of the interaction, and \iac{this} is customary for the consistency of the  Hamiltonian description \iac{of the system} \cite{Todorov_2012_PhysRevB.85.045304, Todorov_few-electron_USC_LC_circuit_PhysRevX.4.041031,DeBernardis_cavityQED_nonperturbative_PhysRevA.97.043820, Rokaj_2018_no_groundstate,rubio_P2_ACSPh_Relevance_of_the_Quadratic_Diamagnetic}.

Interestingly, when the dressing field exhibits a substantial light-matter coupling, it can alter the properties of the electromagnetic field without directly affecting the emitter. 
\iac{As a result}, the emitter now experiences a modified electromagnetic environment, which affects its vacuum properties \iac{such as} its internal electromagnetic interactions, its spontaneous emission rate and, \iac{possibly}, its non-linear properties, as \iac{we are going to see in detail} in the next sections.

\subsection{\ddb{Effective dresser Hamiltonian and} dissipations in \iac{the} dipole gauge picture}

A good way to understand the impact of the \iac{modified} electromagnetic environment \iac{onto the emitter} is to trace out the dressed electromagnetic field and derive an effective description of the emitter only.
\ddb{Since there is no direct coupling between the emitter and the dresser, we can interpret the emitter-cavity coupling $\mathbf{D}\cdot \mathbf{P}_e$ as a linear interaction between the emitter and a generic bath described by the electric displacement field operator.}
Under general assumptions of weak coupling and Markovianity 
%Assuming that the emitter is not ultrastrongly coupled to the electromagnetic environment, we can employ perturbative arguments to describe its interaction with light. The interaction Hamiltonian is given by
%\begin{equation}\label{eq:ham_emitter_D_general}
%H_{e,\,I} = - \int d^3 r~  \frac{\textbf{D}(\textbf{r})\cdot \textbf{P}_{e}(\textbf{r})}{\epsilon_0} + \int d^3 r ~ \frac{\textbf{P}^2_{e}(\textbf{r})}{2 \epsilon_0}.
%\end{equation}
we can employ the standard quantum optical master equation derivation to obtain a description only in terms of the electric displacement field correlation function \cite{Breuer_Petruccione_2002theory}, 
\begin{equation}
    C_{ij}(\textbf{r}, \textbf{r}', \omega ) = \frac{1}{i\epsilon_0\hbar}\int_0^{+\infty} dt\, e^{i\omega t} \langle{\rm vac} | D_i (\textbf{r}, t) D_j (\textbf{r}', 0) |{\rm vac}\rangle
    \label{eq:Greenij}
\end{equation}
Here, the indices $i,j = x,y,z$ label the spatial dimensions and the average is taken over the vacuum state, so that $C_{ij}(\textbf{r}, \textbf{r}', \omega)=0$ for negative $\omega<0$.
%the cavity-dresser equilibrium state at temperature $T$.
Notice that, for the sake of simplicity, in the present discussion we assume that the emitter's polarization  density oscillates around a single frequency $\omega>0$. This assumption \iac{could} be relaxed by allowing for a generic frequency dependence of the operators, \iac{but this is not needed for the present work.}

%In order to have a close expression for the Green's function, the time evolution of the electric displacement field can be calculated \iac{under the closed,} unitary dynamics governed by the polariton Hamiltonian as:
%\begin{equation}
%D_i (\textbf{r}, t) = \exp\left[-iH_{\rm pol}t/\hbar\right]D_i (\textbf{r})\exp\left[ iH_{\rm pol}t/\hbar\right].
%\end{equation}
%Alternatively, if we consider the polariton dynamics to be open and coupled to a dissipative reservoir, the evolution can be described via Lindblad evolution \cite{pupillo_chemestry:10.1063/5.0037412, Finkelstein-Shapiro_PhysRevA.101.042102}, formally written in terms of the time evolution under the action of the Linbladian superoperator
%\begin{equation}
%D_i (\textbf{r}, t) = \exp\left[ \mathcal{L}_{\rm pol}t/\hbar\right] D_i (\textbf{r}).    
%\end{equation}
%Here, the polaritonic Lindbladian superoperator  is defined in the standard way from the cavity-dresser Hamiltonian \iac{supplemented by} cavity and dresser dissipations. For the moment we keep $\mathcal{L}_{\rm pol}$ implicit, but \iac{an explicit form will be given} in the following \iac{Subsections.}

In this framework, the emitter's density matrix evolves according to the following master equation:
\begin{equation}
\hbar \partial_t \rho_e = -i \left[H_{\rm eff}, \rho_e \right] + \mathcal{L}_{\rm loss}(\rho_e) %+ \mathcal{L}_{\rm gain}(\rho_e),
\end{equation}
where the \ddb{correlator} \iac{of the electric displacement field} provides an effective description of \iac{the effect of the polariton environment on the emitter dynamics.} %with photon-dressed mediated interactions. 
\ddb{Similarly to the dyadic formalism used with localized quantum emitters \cite{Gunnar_PhysRevA.66.063810, Asenjo-Garcia_PhysRevA.95.033818, Asenjo-Garcia_PhysRevX.7.031024, genes_PRXQuantum.3.010201,jackson_classical_1999}}, the effective Hamiltonian is given by :
\begin{equation}\label{eq:Heff}
\begin{split}
    H_{\rm eff} & = H_e + \frac{1}{2 \epsilon_0}\int d^3 r \, \textbf{P}^2_{e}(\textbf{r})
    \\
    & + \frac{1}{\epsilon_0}\sum_{ij} \int d^3r \, d^3r' \, P^{(+)}_{e,\,i}(\textbf{r}) \, \text{Re}\left[ C_{ij}(\textbf{r}, \textbf{r}', \omega ) \right]P^{(-)}_{e,\,j}(\textbf{r}'),
\end{split}
\end{equation}
and the associated dissipation %and gain are 
is described by:
\begin{equation}\label{eq:eff_lindblad_loss_emitter}
\begin{split}
   \mathcal{L}_{\rm loss}(\rho) &= \frac{1}{2\epsilon_0} \sum_{ij} \int d^3r \, d^3r' \, \text{Im}\left[ C_{ij}(\textbf{r}, \textbf{r}', \omega ) \right] \times \\ &\times \left( 2P_{e,\,i}^{(-)}(\textbf{r}) \rho P_{e,\,j}^{(+)}(\textbf{r}') - \left\{ P_{e,\,i}^{(-)}(\textbf{r}) P_{e,\,j}^{(+)}(\textbf{r}'), \rho \right\} \right), 
\end{split}
\end{equation}
% \begin{equation}
% \begin{split}
% &\mathcal{L}_{\rm gain}(\rho) = -\frac{1}{2\epsilon_0} \sum_{ij} \int d^3r \, d^3r' \, \text{Im}\left[ G_{ij}^{(-)}(\textbf{r}, \textbf{r}', \omega ) \right]  \times \\ &\times\left( 2P_{e,\,i}^{(+)}(\textbf{r}) \rho P_{e,\,j}^{(-)}(\textbf{r}') - \left\{ P_{e,\,i}^{(+)}(\textbf{r}) P_{e,\,j}^{(-)}(\textbf{r}'), \rho \right\} \right).
% \end{split}
% \end{equation}
Here the $(\pm)$ superscripts stand for the positive/negative frequencies components of the operators % and of the Green's function 
as commonly employed in the open description of USC systems \cite{Cattaneo_2019, Beaudoin_PhysRevA.84.043832, leboite_reviewUSC, nori_USC_review, debernardis2023relaxation, debernardis2024nonperturbative}.
% The positive/negative components of the Green's tensor are given by 
% \begin{eqnarray}
% G_{ij}^{(+)}(\textbf{r}, \textbf{r}', \omega ) &=& \int dt\,\frac{ e^{i\omega t}}{\epsilon_0\hbar} \braket{ D_i^{(+)} (\textbf{r}, t) D_j^{(-)} (\textbf{r}', 0) }_T \\ G_{ij}^{(-)}(\textbf{r}, \textbf{r}', \omega ) &=& \int dt\,\frac{e^{i\omega t}}{\epsilon_0\hbar} \braket{ D_i^{(-)} (\textbf{r}, t) D_j^{(+)} (\textbf{r}', 0) }_T\,,
% \end{eqnarray}
% \iac{and satisfy} $G_{ij}(\textbf{r}, \textbf{r}', \omega ) = G_{ij}^{(+)}(\textbf{r}, \textbf{r}', \omega ) + G_{ij}^{(-)}(\textbf{r}, \textbf{r}', \omega )$.

%\ddb{Considering the case without the dresser and taking $\omega\rightarrow 0$}, from Eq. \eqref{eq:Heff} \ddb{we recover the standard longitudinal electrostatic interaction given by the combination between the $P^2$-term and the effective interaction described by the electric displacement correlator tensor \cite{DeBernardis_cavityQED_nonperturbative_PhysRevA.97.043820}.} 
%the depolarization shift generically affecting this type of systems \cite{ando_electronic_1982} is not a consequence of the $P^2$-term alone, as previously pointed out \cite{Todorov_2010_USC_polariton_Dot_PhysRevLett.105.196402,Todorov_2012_PhysRevB.85.045304}, but \iac{is} rather due to the sum between the $P^2$-term and the effective interaction described by the Green tensor \cite{DeBernardis_cavityQED_nonperturbative_PhysRevA.97.043820}.  

\subsection{\iac{Light emission into} a dressed electromagnetic vacuum}
\label{sec:IIemission}

\iac{Throughout most of this work, we focus on the case} where the polaritonic Hamiltonian $H_{\rm pol}$ is quadratic and fully bosonic as is customary in the literature \cite{Pantazopoulos_PhysRevB.109.L201408, pantazopoulos_unconventional_2024, sentef_doi:10.1126/sciadv.aau6969}. 
In our work, this assumption is due to the fact that we will focus only on the intersubband polariton case, where the dresser is provided by a intersubband transition in a planar quantum well, which is described with great accuracy as a bosonic excitation following a quadratic Hamiltonian \cite{ando_electronic_1982, Todorov_2012_PhysRevB.85.045304}. 
As a consequence, in the diagonal basis, also called the polariton basis \cite{Hopfield_PhysRev.112.1555, Ciuti_PhysRevB.72.115303}, the electric displacement field can be formally written as
\begin{equation}\label{eq:electric_displacement_pol_basis_zp_coefficient}
    D_i(\textbf{r}) = \sum_{k,n} D_{(k,n)}^{\rm zp} \left( {\rm w}_{k,i}(\textbf{r})\, c_{n} + {\rm h.c.} \right),
\end{equation}
where \iac{the $c_{n}$ polariton operators satisfy bosonic commutation rules} $[c_{n}, c_{m}^{\dag}] = \delta_{nm}$, and $D_{(k,n)}^{\rm zp}$ \iac{are} the zero-point amplitudes related to the $n$-th polariton state of the $k$-th electromagnetic mode. Here, the indices $k$ and $n$ should be understood as general array indices rather than integers. In fact, in what follows, we will use $k \mapsto \textbf{k}$ as the in-plane wavevector and $n \mapsto (\textbf{k}, \mathrm{lp/up})$, a doublet containing the in-plane wavevector and the indices referring to the lower or upper polariton. This construction will be clear in the following sections, where the specific example with intersubband polaritons will be worked out.

The orthogonal mode functions $\textbf{w}_k(\textbf{r})$ are solutions of the vectorial Helmholtz equation \cite{power_PhysRevA.25.2473}:
\begin{equation}
\vec{\nabla}\times \vec{\nabla}\times \textbf{w}_k(\textbf{r}) - \frac{\omega_k^2}{c^2} \textbf{w}_k(\textbf{r}) = 0,
\end{equation}
and satisfy the transverse condition $\vec{\nabla}\cdot \textbf{w}_k(\textbf{r}) = 0$. In a confined geometry, they have boundary conditions consistent with the standard electric displacement boundaries \cite{jackson_classical_1999}.  For instance, in a metallic cavity of volume $V$, $\textbf{n}\times \textbf{w}_k(\textbf{r}) \big|_{\delta V}= 0$, where $\textbf{n}$ is the normal vector from the cavity mirrors and $\big|_{\delta V}$ denotes that $\textbf{r}$ is \iac{to be taken} at the boundary of the considered volume space.

The general relation with the quantum electrodynamic vacuum is provided by:
\begin{equation}
\begin{split}
    \sum_{i,j} \int d^3r\, d^3r'\, & {\rm w}_{k,i}^{*}(\textbf{r})\, {\rm w}_{k,j}(\textbf{r}')  \braket{{\rm vac}| D_i(\textbf{r})D_j(\textbf{r}') |{\rm vac}} 
    \\
    &= \sum_n | D_{(k,n)}^{\rm zp} |^2,
\end{split}
\end{equation}
where $|{\rm vac}\rangle$ is the full polaritonic vacuum, satisfying $c_n |{\rm vac}\rangle = 0$ for all $n$.

Including the polariton dissipations $\gamma_{n}$ we obtain
\begin{multline}\label{eq:G_pm}
%\begin{split}
    _{ij}(\textbf{r}, \textbf{r}', \omega ) =\\ =\sum_{k,n} \frac{| D_{(k,n)}^{\rm zp} |^2}{ \epsilon_0 \hbar} \frac{1}{\omega - \omega_{n} - i \frac{\gamma_{n}}{2}} {\rm w}_{k, i}(\textbf{r}){\rm w}_{k,j}^{*}(\textbf{r}'), 
%     \\
%     &G_{ij}^{(-)}(\textbf{r}, \textbf{r}', \omega ) = \\ &= \sum_{k,n} \frac{| D_{(k,n)}^{\rm zp} |^2}{\epsilon_0 \hbar} \frac{n_{T}(\omega_{n} )}{\omega + \omega_{n} + i \frac{\gamma_{n}}{2}} {\rm w}_k^{(i)}(\textbf{r}){\rm w}_k^{(j)*}(\textbf{r}').
% \end{split}
\end{multline}
where %$n_{T}(\omega )$ is the Bose distribution at temperature $T$ and 
$\omega_{n}$ \iac{are} the \iac{eigenfrequencies of the polariton modes resulting from the USC of the dresser with the cavity field. 
%As the polariton modes have positive frequencies $\omega_n>0$, the $G_{ij}^{(-)}$ can not have poles at the positive emitter frequency $\omega$. As such, the gain term ${\mathcal L}_{\rm gain}$ vanishes leaving only a dissipative dynamics ${\mathcal L}_{\rm loss}$ of the emitter related to the imaginary part of $G_{ij}^{(+)}$. 
More in specific, the resonant term in \eqref{eq:G_pm} and the mode functions give information on the polaritonic density of states, while the $| D_{(k,n)}^{\rm zp} |^2$ coefficients quantify the strength of the coupling between the emitter and the polaritons.}

Without the matter component provided by the dresser, polaritons coincide with the cavity field itself, and thus $| D_{(k,n)}^{\rm zp} |^2 = \epsilon_0 \hbar \omega_k  \delta_{kn} / (2V)$, where $\delta_{kn}$ is the Kronecker delta (i.e. the indices $n$ and $k$ become exactly the same index). Here, $V$ is the volume of the cavity, and $\omega_k$ and $\textbf{w}_k(\textbf{r})$ are determined solely by the classical Maxwell equations with the appropriate boundary conditions for the specific cavity considered. 
\ddb{Indeed, if we take the zero-frequency limit $\omega\rightarrow 0$, the cavity correlator in Eq. \eqref{eq:G_pm} becomes by definition the transverse delta $C_{ij}(\textbf{r}, \textbf{r}', \omega ) = -\delta_{\perp}(\textbf{r}, \textbf{r}')/2$ \cite{Cohen-Tannoudji:113864, power_PhysRevA.25.2473, rocio_2022observe}. This term is then summed to the $\mathbf{P}_e^2$-term in effective emitter Hamiltonian of Eq. \eqref{eq:Heff}, acquiring only a contribution from the longitudinal component of the emitter's polarization density $H_{\rm eff} \sim \int d^3 r \mathbf{P}_{e\, \parallel}^2(\mathbf{r})/(2\epsilon_0)$ \cite{DeBernardis_cavityQED_nonperturbative_PhysRevA.97.043820}.}
Although the emitter properties, such as spontaneous emission, can theoretically be modified by reshaping the cavity \iac{and, thus, changing the $\textbf{w}_{k}(\textbf{r})$ mode functions and the corresponding density of states~\cite{Rustomji_PhysRevX.11.021004}}, these modifications are due to the spatial geometry of the boundary conditions, resulting in a different $\delta_{\perp}$ with respect to the free space, and are thus not peculiar to the ultrastrong coupling \iac{regime}.

This situation remains unchanged if the dresser is \iac {in a strong but not ultra-strong coupling regime with the cavity field. In this case,} the counter-rotating terms in the light-matter interaction are \iac{in fact} negligible and the cavity-dresser vacuum is to good approximation the empty vacuum \cite{debernardis2024nonperturbative}. However, the effect of the counter-rotating terms scales quadratically in the light-matter coupling and, when the cavity-dresser coupling enters the ultrastrong regime they can no longer be neglected, resulting in a new non-trivial cavity vacuum \cite{Ciuti_PhysRevB.72.115303,debernardis2024nonperturbative}. In this case, as a consequence, the electric displacement zero-point amplitude \iac{$| D_{(n,k)}^{\rm zp} |^2$ is significantly altered from its bare cavity value, $| D_{(n,k)}^{\rm zp} |^2 \neq \epsilon_0 \hbar \omega_k  \delta_{kn} / (2V)$, giving rise to the exciting phenomena that we are going to see in the next Sections.}

%, mixing the cavity components indexed by $k$ with the polariton components associated with the different index $n$.
%The \iac{rich} consequences of this \iac{physics} will be explored in the following through explicit examples and implementations.

\begin{figure}
    \centering
    \includegraphics[width=\columnwidth]{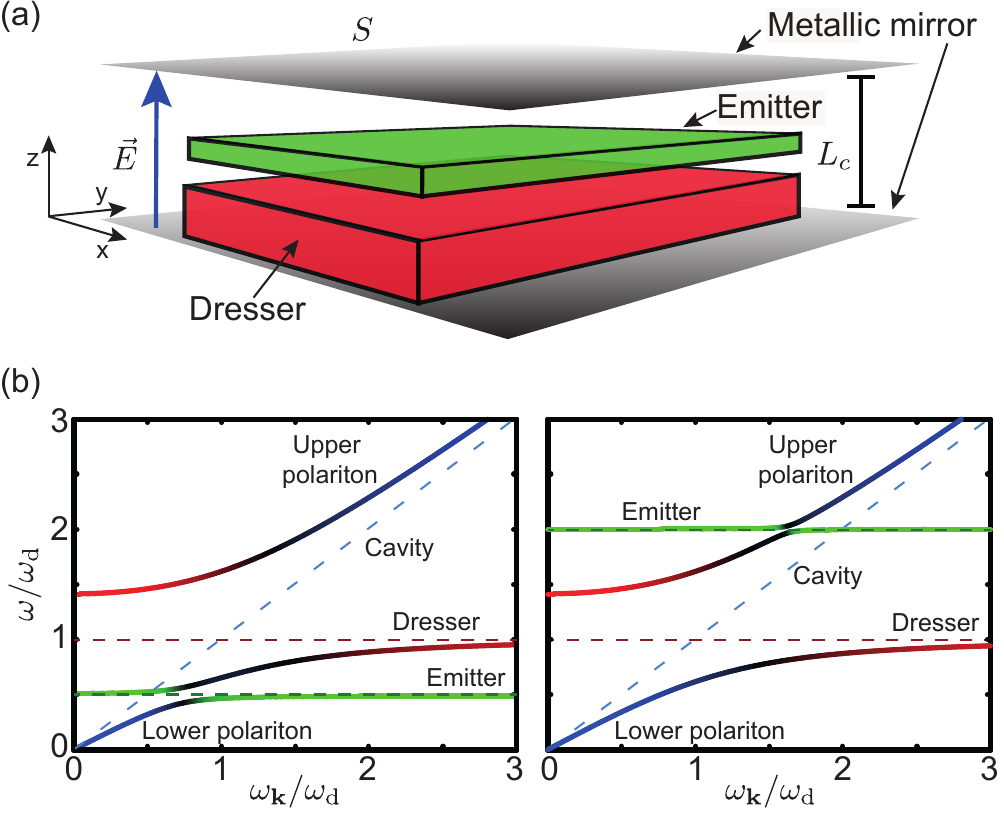}
    \caption{(a) Illustration of the considered setup. The cavity consists of two plane-parallel metallic mirrors of surface $S$ at a distance $L_c$ enclosing a pair of two quantum wells (QWs) called {\em dresser} and {\em emitter}. 
    The electric field of the TM$_0$ cavity modes and the electronic polarization density associated to the intersubband (ISB) transition in the QWs oscillate along a direction perpendicular to the mirrors' plane. (b) Examples of typical polariton spectra emerging from the emitter-cavity-dresser Hamiltonian in Eq. \eqref{eq:ham_plasma_emitter}. The color represents the dominant component %(cavity, dresser, or emitter) 
    of each polaritonic branch: blue, red, green colors respectively refer to the cavity field, dresser, emitter, and black indicates the regions of maximum hybridization. The left and right panels correspond to \iac{the two cases where the emitter is on resonance with either the lower or the upper polariton emerging from the ultra-strong coupling of the dresser and the cavity mode}.
    Parameters: $\Omega_{\rm d}/\omega_{\rm d} = 1$, $\Omega_e/\omega_{\rm d} = 0.1$, \iac{$\omega_e/\omega_{\rm d} = 0.5$ (left panel) or $2$ (right panel).}}
    \label{fig:1}
\end{figure}

\section{Cavity-dresser-emitter with intersubband polaritons}
\label{sec:cavity-dresser_ISB_pol}

\iac{All the discussion of the previous Section was based on very general arguments based on the dipolar gauge description and applies to large class of experimental configurations. In order to obtain concrete predictions to be compared to experiments, starting from this Section we specialize our theory to the most promising case of intersubband polaritons.}

\iac{The specific geometry under investigation is sketched in Fig.~\ref{fig:1}(a): it is based on a planar electromagnetic cavity of surface $S$ along the $xy$ plane and height $L_c$ along the vertical $z$ direction. We focus our attention on the so-called TM$_0$-modes \cite{jackson_classical_1999}: these modes are polarized along the $z$-axis, can have a subwavelength extension along $z$, and are at the heart of semiconductor-based cavity QED setups based on quantum well ISB  transitions~\cite{Todorov_2012_PhysRevB.85.045304, Todorov_few-electron_USC_LC_circuit_PhysRevX.4.041031}.}
%Following our cavity geometry choice, the TM$_0$ cavity electric field results .

\iac{The planar cavity hosts two polarizable QW slabs well separated in space along the direction $z$ perpendicular to the cavity plane. One of them, called the \emph{dresser} is heavily doped so to be in a ultra-strong coupling regime with the cavity mode. 
The dressed vacuum of the USC regime is then used to non-perturbatively influence the coupling to the electromagnetic field of the other QW, called the \emph{emitter}. This latter QW is taken to be much less doped, so that the coupling to light of its own ISB transition  is far below the ultra-strong coupling regime.}
%It is then convenient to differentiate between the two slabs, for which we call \emph{dresser} the most heavily doped one, and \emph{emitter} the other. {\color{red} questo e' gia stato detto}

\subsection{Intersubband polariton Hamiltonian}

Working with intersubband polaritons has the advantage that the excitations of the system can be approximated as harmonic, thereby reducing the complexity of the discussion.
The total Hamiltonian \ddb{describing the interaction between the intersubband quantum wells and the transverse cavity field is}
\begin{equation}\label{eq:ham_plasma_emitter}
\begin{aligned}
    H \approx & H_{c-\text{d}} + \hbar \omega_e \sum_{\textbf{k}} b_{\textbf{k}}^{\dag} b_{\textbf{k}}+ \\
             & -i\frac{\hbar \Omega_e}{2} \sum_{\textbf{k}} \sqrt{ \frac{\omega_{\textbf{k}}}{\omega_{e}} } \left( a_{\textbf{k}} -  a_{-\textbf{k}}^{\dag} \right) \left( b_{-\textbf{k}} + b_{\textbf{k}}^{\dag} \right)+
             \\
             & + \frac{\hbar \Omega_e^2}{4\omega_e} \sum_{\textbf{k}} \left( b_{-\textbf{k}} + b_{\textbf{k}}^{\dag} \right) \left( b_{\textbf{k}} + b_{-\textbf{k}}^{\dag} \right)
\end{aligned}
\end{equation}
\iac{where}
\begin{equation}\label{eq:Ham_c-d}
    \begin{aligned}
        H_{c-\text{d}} &= \hbar \omega_{\text{d}} \sum_{\textbf{k}} d_{\textbf{k}}^{\dag} d_{\textbf{k}} + \sum_{\textbf{k}} \hbar \omega_{\textbf{k}}a_{\textbf{k}}^{\dag} a_{\textbf{k}} \\
        &\quad -i\frac{\hbar \Omega_{\text{d}}}{2} \sum_{\textbf{k}}\sqrt{\frac{\omega_{\textbf{k}}}{\omega_{\text{d}}}} \left(a_{\textbf{k}} -  a_{-\textbf{k}}^{\dag}\right) \left(d_{-\textbf{k}} + d_{\textbf{k}}^{\dag}\right) \\
        &\quad + \frac{\hbar \Omega_{\text{d}}^2}{4\omega_{\text{d}}} \sum_{\textbf{k}} \left(d_{\textbf{k}} + d_{-\textbf{k}}^{\dag}\right) \left(d_{\textbf{k}} + d_{-\textbf{k}}^{\dag}\right),
    \end{aligned}
\end{equation}
describes the (arbitrarily large) coupling of the dresser QW to the cavity.
Here, $a_{\textbf{k}}$ is the annihilation operator of the TM$_0$-cavity photon mode at in-plane wavevector $\textbf{k}$, with dispersion relation $\omega_{\textbf{k}}$. The $b_{\textbf{k}}$ and $d_{\textbf{k}}$ operators are, instead, the annihilation operators of collective ISB excitations of wavevector $\textbf{k}$ in the emitter or dresser QW, with a ${\textbf{k}}$-independent frequency $\omega_{{\rm d},e}$. \iac{As already mentioned,} restricting ourselves to a weak excitation regime, the $b_{\textbf{k}}$ and $d_{\textbf{k}}$ operators can be safely approximated as bosonic~\cite{cominotti2023theory}.

The strength of the light-matter coupling of each QW is quantified by the  dresser and emitter Rabi frequencies $\Omega_{{\rm d},e}$ parameters, which are determined by the corresponding 2D electron densities $n_{{\rm d},e}$ (and thus by the doping density) \iac{via the relation~\cite{Todorov_2012_PhysRevB.85.045304}
\begin{equation}%\label{eq:rabi_freq_emitter_dresser}
    \Omega_{{\rm d},e}^2 =   \frac{f_{{\rm d},e} e^2 n_{{\rm d},e}}{\epsilon_0 m L_c}.
\end{equation}
where} $e$ is the electron charge and $m$ is the effective electron mass. $f_{{\rm d},e}$ is the adimensional oscillator strength parameter determined by the overlap of the electronic wavefunctions in the QW~\cite{Todorov_2012_PhysRevB.85.045304} and exactly equal to $1$ in the case of parabolic wells~\cite{Ciuti_PhysRevB.72.115303}. 
Assuming all other parameters to be constant, the scaling of $\Omega_{{\rm d},e}$ with the doping level $n_{{\rm d},e}$ allows to experimentally control the strength of the light-matter coupling in the dresser and the emitter. The ISB resonance frequencies $\omega_{d,e}$ are then tuned by the geometry and the depth of the confinement potential in the QWs.
\ddb{Importantly, as explained in App. \ref{app:cQED_ham_derivation}, we absorb the contribution deriving from the longitudinal part of the individual $\mathbf{P}_{\rm e,{\rm d}}^2$-term directly into the definition of $\omega_{e/{\rm d}}$. This contribution is crucial for the so called \emph{depolarization shift} \cite{ando_electronic_1982} and takes also account for the effect of the static image charges. However it is irrelevant for our developments and, for the sake of simplicity, we do not explicitly show it. }

\subsection{Emitter-polariton \iac{Rabi splitting}}
%\section{Polariton-Matter interactions}
As already mentioned, in this work we focus on a regime where the emitter is much less doped than the dresser, $n_e \ll n_{\rm d}$, so the coupling of emitter to light $\Omega_{e}$ is much smaller than all other frequencies and can be taken at lowest order while the highly doped dresser is in the USC regime, 
% ,
% \begin{equation}
%     \Omega_e \ll \lbrace{ \omega_e, ~ \omega_{\rm d}, ~ \omega_{k}\rbrace} \simeq \Omega_{\rm d}.
% \end{equation}
$\Omega_e \ll \lbrace{ \omega_e, ~ \omega_{\rm d}, ~ \omega_{\textbf{k}}\rbrace} \simeq \Omega_{\rm d}$. \iac{Under this assumption}, the emitter no longer probes the bare cavity photon modes %, as suggested by the original Hamiltonian \eqref{eq:ham_plasma_emitter}, 
but rather the cavity-dresser polariton modes resulting from the hybridization of the cavity photons and the dresser ISB excitations due to $H_{c-\text{d}}$ in \eqref{eq:Ham_c-d}.

The full polariton spectra arising from the mixing of all three modes shown in Fig.\ref{fig:1}(b) display a selective anticrossing of the emitter (green) mode with the lower (left panel) or the upper (right panel) cavity-dresser polariton depending on the value of the emitter frequency: in the Figure, for each polariton branch, the color indicates the dominant cavity (blue), dresser (red), or emitter (green) nature, and black represents a maximal mixing. As a most remarkable feature, for the same value of $\Omega_e$, we notice that the anti-crossing with the lower polariton is much wider than the one with the upper polariton.

\begin{figure}
    \centering
    \includegraphics[width=\columnwidth]{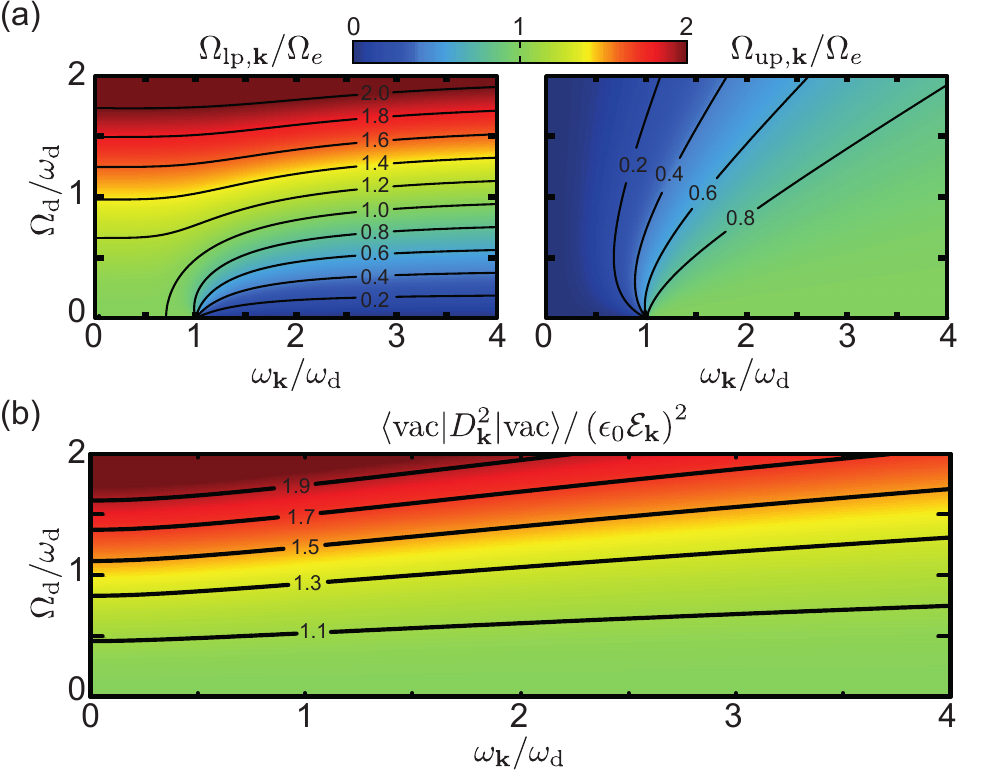}
    \caption{(a) Colorplot of the polariton-vacuum Rabi frequencies as a function of the photon mode frequency $\omega_{\textbf{k}}$
    and of the dresser Rabi frequency $\Omega_{\rm d}$, for an emitter resonant with the lower ($\omega_e = \omega_{{\rm lp}, \textbf{k}}$, left panel) or upper ($\omega_e = \omega_{{\rm up}, \textbf{k}}$, right panel) polariton.
    (b) Colorplot of the vacuum fluctuations of the displacement field in the TM$_0$ mode. }
    \label{fig:2}
\end{figure}

%As we are restricting to the lowest-order terms in the emitter coupling $\Omega_e$, 
\iac{This is a key result of this work. In order to physically understand it,} we can rewrite the total Hamiltonian of Eq.\eqref{eq:ham_plasma_emitter} in the polariton basis within the emitter-polariton rotating-wave approximation
\begin{equation}
\begin{aligned}
    H &\approx \hbar \omega_e \sum_{\textbf{k}} b_{\textbf{k}}^{\dag} b_{\textbf{k}}+ \\
      &\quad + \sum_{\textbf{k}} \hbar \omega_{\text{up}, \textbf{k}} p_{\text{up}, \textbf{k}}^{\dag} p_{\text{up}, \textbf{k}} + \sum_{\textbf{k}} \hbar \omega_{\text{lp}, \textbf{k}} p_{\text{lp}, \textbf{k}}^{\dag} p_{\text{lp}, \textbf{k}}+ \\
      &\quad + i\sum_{\textbf{k}}\left( \frac{\hbar \Omega_{\text{up}, \textbf{k}}}{2} p_{\text{up}, \textbf{k}} b_{\textbf{k}}^{\dag} + \frac{\hbar \Omega_{\text{lp}, \textbf{k}}}{2} p_{\text{lp}, \textbf{k}} b_{\textbf{k}}^{\dag}\right) + \text{h.c.},
\end{aligned}
\label{eq:Hpol}
\end{equation}
where $p_{\, {\rm up}, \textbf{k}}, p_{\, {\rm lp}, \textbf{k}}$ are the annihilation operators of a cavity-dresser polariton in the upper or lower polariton branches of %the dresser-cavity Hamiltonian 
\eqref{eq:Ham_c-d} with eigenfrequencies \begin{equation}
    \omega_{{\rm up/lp}, \textbf{k}}^2 = \frac{\omega^2_{ \textbf{k}} + \bar{\omega}_{\rm d}^2 %+ \Omega_{\rm d}^2
    }{2} \pm \sqrt{\frac{\left( \bar{\omega}_{\rm d}^2 %+ \Omega_{\rm d}^2 
    - \omega^2_{ \textbf{k}} \right)^2}{4} + \Omega_{\rm d}^2\omega^2_{ \textbf{k}}}.
\end{equation}%, as labeled in Fig. \ref{fig:1}(b).
%As it is discussed in full detail in Secs.\ref{app:polariton_canonical_transform} and \ref{app:polaritonic_photon} of the SM, 
where $\bar{\omega}_{\rm d}=\sqrt{\omega^2_{\rm d}+\Omega^2_{\rm d}}$ includes the quadratic shift of the dresser frequency associated to the polariton gap. %~\cite{zaluzny1982inter,ando_electronic_1982,cominotti2023theory}.
The effective polariton-vacuum Rabi frequencies quantifying the coupling strength between the emitter and the lower/upper cavity-dresser polaritons read (see App. \ref{app:polariton_canonical_transform}-\ref{app:polaritonic_photon})
%\ref{app:polariton_canonical_transform} and \ref{app:polaritonic_photon})
\begin{equation}\label{eq:Rabi_pol}
%\begin{split}
\frac{\Omega_{{\rm up}, \textbf{k}}}{\Omega_e}  =  \sqrt{\frac{ \omega^2_{\textbf{k} } }{\omega_e \omega_{{\rm up}, \textbf{k}}}} \sin \theta_{\textbf{k}} \textrm{~,~~}
\frac{\Omega_{{\rm lp}, \textbf{k}}}{\Omega_e} =  \sqrt{\frac{ \omega^2_{\textbf{k} } }{\omega_e \omega_{{\rm lp}, \textbf{k}}}} \cos \theta_{\textbf{k}}\, ,%\\
%\end{split}
\end{equation}
\iac{where the hybridization angle $\theta_{\textbf{k}}$ is defines as}
\begin{equation}
    %\begin{array}{cc}
         \cos^2 \theta_{\textbf{k}} = \frac{\omega_{{\rm up}, \textbf{k}}^2 - \omega_{\textbf{k}}^2}{\omega_{{\rm up}, \textbf{k}}^2 - \omega_{{\rm lp}, \textbf{k}}^2} \textrm{~,~~} \sin^2 \theta_{\textbf{k}} = \frac{ \omega_{\textbf{k}}^2 - \omega_{{\rm lp}, \textbf{k}}^2}{\omega_{{\rm up}, \textbf{k}}^2 - \omega_{{\rm lp}, \textbf{k}}^2}.
%    \end{array}
\end{equation}
Interestingly, all the information regarding the hybridization between the cavity photon and the dresser excitations due to the USC is contained in the hybridization angle $\theta_{\textbf{k}}$, which 
%. This quantity allows us to summarize 
summarizes into a single parameter the Hopfield coefficients expressing the polariton operators $p_{\text{lp}, \textbf{k}}$ and $p_{\text{up}, \textbf{k}}$ in terms of the cavity photon $a_{\textbf{k}}$ and dresser $d_{\textbf{k}}$ operators and their hermitian conjugates~\cite{Ciuti_PhysRevB.72.115303,deLiberato_virtualphotons}.

The reformulation in terms of the polariton Hamiltonian \eqref{eq:Hpol} provides a physical understanding of the peculiar features displayed by the emitter-polariton coupling that have been observed in Fig.\ref{fig:1}(b). In Fig. \ref{fig:2}(a) we show a color plot of the polariton-vacuum Rabi frequencies $\Omega_{{\rm lp}, \textbf{k}}$ and $\Omega_{{\rm up},\textbf{k}}$ as a function of the wavenumber $\textbf{k}$ and of the dresser Rabi frequency $\Omega_{\rm d}$, when the emitter is resonant with some state on the lower $\omega_e = \omega_{{\rm lp}, \textbf{k}}$ (left panel) or the upper $\omega_e = \omega_{{\rm up}, \textbf{k}}$ (right panel) polariton branch. In the full polariton spectrum of Fig.\ref{fig:1}(b), $\Omega_{{\rm lp/up}, \textbf{k}}$ quantifies the magnitude of the Rabi splitting.

For a weak dresser Rabi frequency  $\Omega_{\rm d} \ll \omega_{\rm d}$, the resonant lower and upper polariton-vacuum Rabi frequencies display similar behavior, with $\Omega_{{\rm lp (up)}, \textbf{k}} \simeq \Omega_e$ when the polariton has a fully photonic nature and, \iac{instead}, $\Omega_{{\rm up (lp)}, \textbf{k}}\simeq 0$ when the polariton has a fully excitonic nature. In general, while the photonic weight is redistributed between the upper and lower polaritons, the total weight is conserved, $\Omega_{{\rm lp}, \textbf{k}}^2+\Omega_{{\rm up}, \textbf{k}}^2\approx\Omega_e^2$, as a consequence of the weakly dressed regime where $\omega_{{\rm up/lp}, \textbf{k}} \simeq \omega_{\textbf{k}}$.
% (except in the small region around the cavity-dresser anticrossing).

The physics drastically changes when the dresser enters the USC regime for $\Omega_{\rm d} \simeq \omega_{\rm d}$. Here $\omega_{{\rm up/lp}, \textbf{k}} \neq \omega_{\textbf{k}}$, and the square-root prefactors %s{\color{red} $\left({ \omega_{\textbf{k} } }/{\sqrt{\omega_e \omega_{{\rm up/lp}, \textbf{k}}}}\right)$}
in \eqref{eq:Rabi_pol} start to matter: the coupling with the lower polariton $\Omega_{{\rm lp}, \textbf{k}}$ is reinforced and remains significant up to higher wavenumbers, while the upper polariton's coupling $\Omega_{{\rm up}, \textbf{k}}$ is significantly reduced. As an immediate consequence the conservation of the photonic weight is strongly violated, $\Omega_{{\rm lp}, \textbf{k}}^2+\Omega_{{\rm up}, \textbf{k}}^2 \neq \Omega_e^2$.

This quite remarkable behavior is due to the mixing of creation and annihilation operators in the Bogoliubov transformation to polariton operators~\cite{Ciuti_PhysRevB.72.115303}, so that the normal and anomalous terms constructively (destructively) interfere in determining the \iac{effective strength of the coupling of the emitter to the lower (upper) polariton}.
%\emph{Experimental observability}--- 
\iac{As it is discussed in App. \ref{app:detailes_protocol}, this} marked asymmetry of the Rabi splitting of the lower and upper polariton branches is straightforwardly observed in a cavity transmission/reflection spectroscopy experiment.
%Fig.\ref{fig:4} displays examples of simulated transmission spectra well in the strong emitter-polariton coupling regime $\Omega_{\rm lp/up, \textbf{k}}\gg \gamma_{\rm lp/up}$ where all polariton branches are well separated.

\section{Quantum fluctuations in the USC vacuum}
\label{sec:quantum_fluct_USC_vacuum}

\subsection{Vacuum-field Rabi splitting}
In order to obtain a deeper understanding of the relation between the modified emitter light-matter coupling strength and the properties of the USC dressed vacuum \iac{discussed in the previous Section, we make use of the polariton Hamiltonian to explicitly} evaluate the \iac{quantum} fluctuations of the cavity electric displacement field 
\begin{equation}
D_{\textbf{k}} = i \sqrt{\frac{\epsilon_0 \hbar\omega_{\textbf{k}}}{2SL_c}}(a_{\textbf{k}} -  a_{-\textbf{k}}^{\dag}).
\end{equation}
Within our dipole representation, this field --rather than the electric field-- represents in fact the correct electromagnetic degree of freedom to describe light-matter interactions~\cite{Cohen-Tannoudji:113864, craig1998molecular, Stengel_nat2009_DFT_displacement_field}. 
By using the hybridization angle $\theta_{\textbf{k}}$ defined above, we can express this quantity as (see App. \ref{app:polaritonic_photon})
%\ref{app:hybridization_angle_vac_obs} of the SM)
\iac{\begin{equation}\label{eq:displacement_polariton_basis1}
\begin{split}
    &D(\textbf r) = i \sqrt{\frac{\epsilon_0 \hbar}{2SL_c}} \sum_{\textbf k} \omega_{\textbf k} \, e^{i \textbf k \cdot \textbf{r}_{\parallel}} 
    \\
    &\times \left( \frac{\sin \theta_{\textbf k}}{\sqrt{\omega_{\rm up, \textbf k}}} p_{\,{\rm up}, \textbf k} + \frac{\cos \theta_{\textbf k}}{\sqrt{\omega_{\rm lp, \textbf k}}} p_{\,{\rm lp}, \textbf k} \right) + {\rm h.c.}.
\end{split}
\end{equation}
This straightforwardly leads to}
\begin{multline}
    \frac{\langle {\rm vac}|D^2_{\textbf{k}} |{\rm vac}\rangle}{(\epsilon_0\mathcal{E}_{\textbf{k}})^2} = %\left( 
    \frac{\omega_{\textbf{k}}}{\omega_{\rm up, \textbf{k}}}\sin^2 \theta_{\textbf{k}} + \frac{\omega_{\textbf{k}}}{\omega_{\rm lp, \textbf{k}}}\cos^2 \theta_{\textbf{k}}%\right) 
    \\ = \frac{\omega_e}{\omega_{\textbf{k}}}\left(\frac{\Omega_{\rm lp, \textbf{k}}^2}{\Omega_e^2}+\frac{\Omega_{\rm up, \textbf{k}}^2}{\Omega_e^2}\right),
\label{eq:vacuum}
\end{multline}
where $\mathcal{E}_{\textbf{k}}^2 = (\hbar \omega_{\textbf{k}})/(2\epsilon_0 S L_c)$ are the quantum fluctuations of the electric (or, in this case equivalently, of the displacement) field in a bare cavity.

The peculiarity of the USC is then visible in Fig. \ref{fig:2}(b),
where we display a color plot of the total electric displacement fluctuations in the different $\textbf{k}$ modes as a function of the strength of the cavity dresser coupling $\Omega_d$. 
On the one hand, for weak or moderate $\Omega_d$, the prefactors $\omega_{\textbf{k}}/\omega_{\{\textrm{up,lp}\},\textbf{k}}$ on the first line in Eq. \eqref{eq:vacuum} are close to one and thus play a minor role; thanks to the trigonometric identity $\sin^2\theta_{\textbf{k}}+\cos^2\theta_{\textbf{k}}=1$ associated to the conservation of the photonic weight, the two contributions then sum up to the standard bare vacuum fluctuations.
On the other hand, the total fluctuations get substantially increased in the USC regime, in connection with the increased value of the lower polariton-vacuum Rabi frequencies.

More in specific, the second line of  \eqref{eq:vacuum} shows that the contributions of the lower and upper polariton frequencies to the vacuum fluctuations have an amplitude proportional to the polariton-emitter Rabi frequencies $\Omega_{\rm lp/up, \textbf{k}}$.
%, a result that extends to the USC vacuum case the traditional concept of vacuum-field Rabi splitting~\cite{Mondragon_PhysRevLett.51.550,agarwal_PhysRevLett.53.1732}.
\iac{On the one hand, this provides a bridge to 
the contributions to the zero-point fluctuations coming from the upper and lower polaritons in Eqs.(\ref{eq:electric_displacement_pol_basis_zp_coefficient}-\ref{eq:G_pm}) as }
\begin{align}
    |D_{(\textbf{k}, {\rm up})}^{\rm zp}|^2 = \frac{\epsilon_0\hbar\omega_e}{2 S L_c} \frac{\Omega_{\rm up, \textbf{k}}^2}{\Omega_e^2} \, , 
    &&
    |D_{(\textbf{k}, {\rm lp})}^{\rm zp}|^2 = \frac{\epsilon_0\hbar\omega_e}{2 S L_c} \frac{\Omega_{\rm lp, \textbf{k}}^2}{\Omega_e^2}.
\end{align}
\iac{On the other hand, the same equation extends to the USC vacuum case the traditional concept of vacuum-field Rabi splitting, whose magnitude is indeed determined by the strength of the quantum fluctuations~\cite{Mondragon_PhysRevLett.51.550,agarwal_PhysRevLett.53.1732}.}

\subsection{Polariton modified \iac{dipole-dipole} %electron-electron
interactions}

While the Rabi splitting in the strong emitter-polariton coupling regime provides a most direct experimental signature of the modified Rabi couplings \eqref{eq:Rabi_pol}, it is worth to explore the consequences that the modified polariton vacuum has on the \iac{dynamics of the emitter} itself.
In order to do so, we focus on the case when the emitter is weakly coupled to the polaritons and we can apply the \ddb{master equation} formalism illustrated in Sec. \ref{sec:general_framework}, \ddb{where the electric displacement correlator is projected on the TM$_0$ modes only}:
\begin{equation}
\begin{split}\label{eq:G_ISBpol_TM0}
     C_{{\rm TM}_0}&(\textbf{r}, \textbf{r}', \omega ) = \frac{1}{2 }\frac{\omega_e}{\Omega_e^2} \int \frac{d^2k}{(2\pi)^2}e^{i\textbf{k}\cdot (\textbf{r} - \textbf{r}')} \, \times
    \\
    &   \Bigg[ \frac{\Omega_{\rm lp, \textbf{k}}^2}{\omega - \omega_{\text{lp}, \textbf{k}} - i\gamma_{\text{lp}, \textbf{k}}/2 } + \frac{\Omega_{\rm up, \textbf{k}}^2}{\omega - \omega_{\text{up}, \textbf{k}} - i\gamma_{\text{up}, \textbf{k}}/2 }\Bigg].
\end{split}
\end{equation}
The restriction to \iac{the TM$_0$ cavity modes} \ddb{is well motivated and experimentally tested} for the setups under our investigation \cite{Todorov_2012_PhysRevB.85.045304, Todorov_2010_USC_polariton_Dot_PhysRevLett.105.196402, Todorov_few-electron_USC_LC_circuit_PhysRevX.4.041031, tredicucci_PhysRevB.79.201303, Colombelli_immunity_PhysRevB.96.235301, Colombelli_Perspectives_PhysRevX.5.011031}, thus giving the largest contribution to the cavity mediated effective interactions and losses.
In this way the modifications of the correlator due to the presence of the ISB dresser will affect only its component given by the TM$_0$ modes, justifying the definition given in Eq. \eqref{eq:G_ISBpol_TM0}. \iac{While an accurate description of the other components may be in general important to get quantitative predictions, they are not substantially affected by the modified quantum vacuum and their main modifications stem from the modified boundary conditions in the cavity mode functions $\textbf{w}$ described in Sec.\ref{sec:general_framework} and in App.\ref{app:cQED_ham_derivation}: 
as remarked above we label these changes as purely geometrical and not due to the presence of the dressing material, so, for clarity, we do not specifically consider them in this work.}

The behavior of the polaritonic correlation function  \eqref{eq:G_ISBpol_TM0} is now strongly inhomogeneous in $\omega$ and depends on the polaritonic losses too. Its complete characterization is not possible here, so we decided to focus \iac{on a few most relevant cases only:}

%\begin{figure}
%    \centering
%    \includegraphics[width=\columnwidth]{fig4.pdf}
%    \caption{Colorplot of the transmission spectrum as a function of the incident wavevector $\textbf{k}$ and frequency $\omega$. The white dashed line marks the wavevector location of the emitter minimal anticrossing, the green dashed line marks the emitter frequency $\omega_e$. The two yellow circles mark the locations where the polariton maximally hybridizes with the emitter. As in the previous figures, the left/right panels correspond to an emitter on resonance with the lower/upper polariton. 
%    Parameters: $\Omega_e/\omega_{\rm d} = 0.2$, 
%    $\Omega_{\rm d}/\omega_{\rm d}= 1$, $\omega_e/\omega_{\rm d} = 0.7$ (left panel) and $1.6$ (right panel). The photonic, dresser and emitter decay rates are $\gamma/\omega_{\rm d} = 0.01$, $\kappa_{\rm d} = \kappa_{e} = 0.05\, \omega_{\rm d}$.}
%    \label{fig:4}
%\end{figure}

\begin{enumerate}
    \item Electrostatic (or adiabatic) limit, when the emitter's frequency is much slower than the typical polaritonic time scales and it can be taken as vanishing $\omega_e\rightarrow 0$. This case is particularly relevant for the current discussion related to possible modifications of material properties such that superconductivity \cite{andolina_PhysRevB.102.125137}.
    \item Band-gap limit, when the emitter's frequency lays in the polaritonic band-gap, $\omega_e \in [\omega_{\rm d}:\omega_{\rm d} + \Omega_{\rm d}^2/\omega_{\rm d}]$.
\end{enumerate}
In both cases we assume negligible losses $\gamma_{\rm up/lp}\approx 0$, and we assume that there are no poles, since $\omega\neq \omega_{\text{up}/\text{lp}, \textbf{k}}$.
We thus obtain the compact expression
\begin{equation}
\begin{split}
    C_{{\rm TM}_0}&(\textbf{r}, \textbf{r}', \omega ) = \frac{1}{2}\int \frac{d^2k}{(2\pi)^2}e^{i\textbf{k}\cdot (\textbf{r} - \textbf{r}')} \times
    \\
    & \left[ \frac{\omega_{\textbf{k}}^2}{\omega_{\text{lp}, \textbf{k}}^2 }\cos^2 \theta_{\textbf{k}} \frac{\omega_{\text{lp}, \textbf{k}}}{\omega - \omega_{\text{lp}, \textbf{k}}}+ \frac{\omega_{\textbf{k}}^2}{\omega_{\text{up}, \textbf{k}}^2}\sin^2 \theta_{\textbf{k}}\frac{\omega_{\text{up}, \textbf{k}}}{\omega - \omega_{\text{up}, \textbf{k}}}\right].
\end{split}
\end{equation}

Interestingly using the various relations between the polariton eigenfrequencies and their USC mixing angle developed in App. \ref{app:polariton_canonical_transform} we have the identiy
\begin{equation}
    \frac{\omega_{\textbf{k}}^2}{\omega_{\text{lp}, \textbf{k}}^2 }\cos^2 \theta_{\textbf{k}} + \frac{\omega_{\textbf{k}}^2}{\omega_{\text{up}, \textbf{k}}^2}\sin^2 \theta_{\textbf{k}} =  1 + \frac{\Omega_{\rm d}^2}{\omega_{\rm d}^2}.
\end{equation}
Introducing the Fourier kernel
\begin{equation}
\begin{split}
    K_{\omega}(\textbf{k} )  = &\frac{\omega_{\textbf{k}}^2}{\omega_{\text{lp}, \textbf{k}}^2 }\cos^2 \theta_{\textbf{k}} \frac{\omega}{\omega - \omega_{\text{lp}, \textbf{k}}} - \frac{\Omega_{\rm d}^2}{\omega_{\rm d}^2}\frac{\omega}{\omega-\omega_{\rm d}} 
    \\
    & + \frac{\omega_{\textbf{k}}^2}{\omega_{\text{up}, \textbf{k}}^2}\sin^2 \theta_{\textbf{k}}\frac{\omega}{\omega - \omega_{\text{up}, \textbf{k}}} 
\end{split}
\end{equation}
\iac{we can} re-write the correlation function as
\begin{equation}\label{eq:Geff_tm0_explicit}
    \begin{split}
        C_{{\rm TM}_0}(\textbf{r}, \textbf{r}', \omega ) &= -\frac{1}{2}\delta(\textbf{r} - \textbf{r}') -\frac{1}{2}\frac{\Omega_{\rm d}^2}{\omega_{\rm d}^2} \frac{\omega_{\rm d}}{\omega_{\rm d}-\omega} \delta(\textbf{r} - \textbf{r}')
        \\
        & + \frac{1}{2}\int \frac{d^2k}{(2\pi)^2} \, K_{\omega}(\textbf{k} ) \, e^{i\textbf{k}\cdot (\textbf{r} - \textbf{r}')},
    \end{split}
\end{equation}
\iac{Plugging this expression into the emitter's effective Hamiltonian in Eq. \eqref{eq:Heff} together with the representation given in App. \ref{app:cQED_ham_derivation} for the ISB emitter, we obtain}
\begin{equation}\label{eq:ham_emitter_isb_eff_int}
\begin{split}
    H_{\rm eff} &\approx  \hbar \omega_e \sum_{\textbf{k}}b^{\dag}_{\textbf{k}} b_{\textbf{k}} 
    \\
    & - \frac{\hbar \Omega_e^2}{4\omega_e\omega_{\rm d}} \frac{\Omega_{\rm d}^2}{\omega_{\rm d}-\omega_e} \sum_{\textbf{k}} \left( b_{-\textbf{k}} + b_{\textbf{k}}^{\dag} \right) \left( b_{\textbf{k}} + b_{-\textbf{k}}^{\dag} \right)
    \\
    & + \frac{\hbar \Omega_e^2}{4\omega_e}\sum_{\textbf{k}} K_{\omega_e}(\textbf{k} )  \left( b_{-\textbf{k}} + b_{\textbf{k}}^{\dag} \right) \left( b_{\textbf{k}} + b_{-\textbf{k}}^{\dag} \right)
\end{split}
\end{equation}

We immediately see that the first term of the effective interaction in Eq. \eqref{eq:Geff_tm0_explicit} exactly cancels the $\textbf{P}^2$ term. 
%A common misconception is that the $\textbf{P}^2$ term contains the dipole-dipole interaction \ddb{responsible for the ISB depolarization shift} \cite{ando_electronic_1982,Todorov_2012_PhysRevB.85.045304, Todorov_few-electron_USC_LC_circuit_PhysRevX.4.041031}. However, we can see from here that this is not correct as this term drops out from the effective Hamiltonian. 
As a consequence, the physical meaning of this term cannot be separated from the $\textbf{D}\cdot \textbf{P}$ term, and they should always be interpreted together \cite{DeBernardis_cavityQED_nonperturbative_PhysRevA.97.043820, Rokaj_2018_no_groundstate, rubio_P2_ACSPh_Relevance_of_the_Quadratic_Diamagnetic}.
The other two terms do not experience any cancellations and thus represent the physical interaction mediated by the polaritonic electromagnetic field in the electrostatic and band-gap limit. \ddb{Their immediate physical consequence in a ISB experiment is well visible from the emitter's effective Hamiltonian in Eq. \eqref{eq:ham_emitter_isb_eff_int}: they represent a new contribution to the emitter's depolarization shift due to the presence of the dresser.}

\ddb{In particular,} the second term of Eq. \eqref{eq:Geff_tm0_explicit} is the bare electrostatic contribution, representing the interaction with the dresser's dipole electric field in a planar slab geometry. It scales quadratically with the emitter and dresser coupling strengths, $\sim O(\Omega_e^2\Omega_{\rm d}^2)$. Notice that in the electrostatic limit, $\omega_e = 0$, it comes with a minus sign, indicating that the energy of the emitter is lowered in the presence of the dresser (like in presence of an attractive force), while it can become positive (an thus increasing the emitter's energy like in presence of a repulsive force) when the emitter frequency is in the band-gap, $\omega_e \in [\omega_{\rm d}:\omega_{\rm d} + \Omega_{\rm d}^2/\omega_{\rm d}]$.
%It must not be confused with the single-slab depolarization shift \cite{ando_electronic_1982}, which has a positive sign, and is typically the dominant contribution \cite{Geiser_PhysRevLett.108.106402, Shtrichman_depshift_PhysRevB.65.035310, Muravev_PhysRevB.93.041110} coming from the electron-electron Coulomb interaction within each slab.

\begin{figure}
    \centering
    \includegraphics[width=\columnwidth]{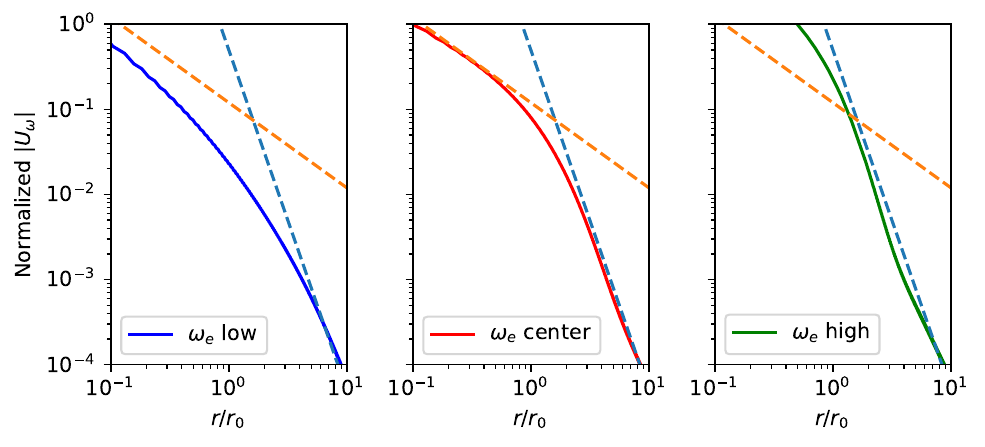}
    \caption{Finite frequency residual effective potential $U_{\omega}(\textbf{r})$ as a function of the distance $r$. In all three plots the potential is normalized \iac{to its the maximum value in} the central panel, the distance is expressed in units of $r_0 = c/\omega_{\rm d}$ and we used $\Omega_{\rm d}/\omega_{\rm d} = 1$. In the left panel $\omega_e = (\sqrt{\omega_{\rm d}^2 + \Omega_{\rm d}^2} - \omega_{\rm d})/5 + \omega_{\rm d}/2$ is closer to the lower bandgap-edge, in the central panel $\omega_e = (\sqrt{\omega_{\rm d}^2 + \Omega_{\rm d}^2} - \omega_{\rm d})/2 + \omega_{\rm d}/2$ is exactly at the center of the bandgap, in the right panel $\omega_e = (\sqrt{\omega_{\rm d}^2 + \Omega_{\rm d}^2} - \omega_{\rm d})/1.1 + \omega_{\rm d}/2$ is closer to the upper bandgap-edge.
    The dashed lines indicates the scaling $\sim 1/r$ (orange) and $1/r^4$ (blue). The Fourier transform \eqref{eq:FFT} is performed numerically via fft.  }
    \label{fig:3}
\end{figure}

The third term of Eq. \eqref{eq:Geff_tm0_explicit} represents a finite frequency correction due to retardation, falling into the category of Casimir-Polder forces \cite{Casimir_polder_origina_PhysRev.73.360, CasimirPolder_review_RevModPhys.81.1827}. We found an approximate linear scaling with the dresser coupling $\sim O(\Omega_e^2\Omega_{\rm d})$. It gives a negative contribution to the emitter's energy $K_{\omega_e}(\textbf{k}) \leq 0$ and is non-zero only when the emitter is in the band-gap.
Differently from the other term, which is strictly a contact interaction term, this one has a finite range of interaction, which changes dependently on the emitter's frequency and from the dresser light-matter coupling.
In Fig. \ref{fig:3} we show three examples of the resulting finite-range potential \iac{\begin{equation}U_{\omega}(\textbf{r}) \sim \int \frac{d^2k}{(2\pi)^2} \, K_{\omega}(\textbf{k} ) \, e^{i\textbf{k}\cdot (\textbf{r} - \textbf{r}')}
\label{eq:FFT}
\end{equation}
in the three cases $\omega_{e} \simeq \omega_{\rm d}^+$ (left) $\omega_{e} \simeq (\sqrt{\omega_{\rm d}^2 + \Omega_{\rm d}^2} - \omega_{\rm d})/2 + \omega_{\rm d}/2 $ (center), $\omega_{e} \simeq (\sqrt{\omega_{\rm d}^2 + \Omega_{\rm d}^2})^-$ (right), while keeping $\Omega_{\rm d}/\omega_{\rm d} = 1$.}

Due to their quadratic scaling in the emitter and dresser vacuum Rabi frequency these terms are obviously very small in any system subjected only to the strong coupling regime.
However in ultrastrongly coupled \iac{systems}, where the dresser vacuum Rabi frequency $\Omega_{\rm d}$ is larger than any other energy scale  \cite{nori_USC_review, mueller_deep_2020}, these terms provides a new source of interactions that can be exploited, for instance, to boost or modify polariton interactions \cite{bamba_PhysRevB.87.235322, carusotto_RevModPhys.85.299, cominotti2023theory}.

\subsection{Spontaneous emission}

\iac{As it was anticipated in Sec.\ref{sec:general_framework}, the coupling to a modified} polaritonic vacuum may have also strong consequences on the spontaneous emission rate in the weak emitter-polariton coupling regime 
%where the polariton linewidth exceeds the polariton-vacuum Rabi frequencies 
$\gamma_{{\rm lp/up}, \textbf{k}} \gg \Omega_{{\rm lp/up}, \textbf{k}}$ where the Rabi oscillations are replaced by an irreversible emission process. 

\iac{The spontaneous emission rate can be estimated by combining the correlation function in Eq.\eqref{eq:G_ISBpol_TM0} with the prescriptions of Sec.\ref{sec:IIemission}. Given the symmetry of our set-up, in-plane momentum is conserved during the spontaneous emission process. We also assume that the emitter is resonant with a mode of one polariton branch only, while the other polariton branch is completely out of resonance.} 

Due to the finite amount of photon dissipations, \iac{this straightforwardly gives}
\begin{equation}
\begin{split}
    {\rm Im}\left[ \tilde{C}_{{\rm TM}_0}(\textbf{k}, \omega_e = \omega_{\text{lp}, \textbf{k}} ) \right]  \approx \frac{1}{2}\frac{\omega_{\textbf{k}}^2}{\omega_{\text{lp}, \textbf{k}} \gamma_{\text{lp}, \textbf{k}} }\cos^2 \theta_{\textbf{k}}, 
\end{split}
\end{equation}
\begin{equation}
\begin{split}
    {\rm Im}\left[ \tilde{C}_{{\rm TM}_0}(\textbf{k}, \omega_e = \omega_{\text{up}, \textbf{k}} ) \right]  \approx \frac{1}{2}\frac{\omega_{\textbf{k}}^2}{\omega_{\text{up}, \textbf{k}} \gamma_{\text{up}, \textbf{k}} }\sin^2 \theta_{\textbf{k}}.
\end{split}
\end{equation}
where $\tilde{C}_{{\rm TM}_0}$ represents the \iac{spatial} Fourier transform of the correlation function \eqref{eq:G_ISBpol_TM0} in the in-plane wavevector $\textbf{k}$-space.

As it was pointed out in Refs.\cite{Bamba_Maxwell_dissipations_PhysRevA.88.013814,deliberato_comment_to_bamba_rates_PhysRevA.89.017801,DeLiberato_PhysRevLett.112.016401}, the polaritonic linewidths $\gamma_{\text{lp/up}, \textbf{k}}$ have also a dependence from the mixing angle $\theta_{\textbf{k}}$ in USC.
\iac{As a complete description of this physics goes beyond the scope of this work, we restrict here to} the simplest examples where the polaritons dissipate through an Ohmic bath within a \iac{rotating wave approximation for the weak system-bath coupling} (see App. \ref{app:losses_dissipations}).
\iac{In order to obtain compact formulas we focus on the two extreme cases  of cavity-dominated losses and of  dresser-dominated losses.}

In the first case, we have (see App. \ref{app:losses_dissipations})
\begin{eqnarray}
    \gamma_{\text{lp}, \textbf{k}} &=& \gamma \cos^2 \theta_{\textbf{k}}, \\
    \gamma_{\text{up}, \textbf{k}} &=& \gamma \sin^2 \theta_{\textbf{k}}.
\end{eqnarray}
Here $\gamma$ is the bare decay rate of the uncoupled cavity. 
The cavity-dominated spontaneous emission rate mediated by the resonant polariton mode is then immediately extracted from Eq. \eqref{eq:eff_lindblad_loss_emitter} and it reads~\cite{purcell_PhysRev.69.37} 
\begin{eqnarray}
\label{eq:Gamma_gamma_dominated_lp}
    \Gamma_{\rm lp, \textbf{k}}^{\gamma}&\approx& \frac{\Omega_{\rm lp, \textbf{k}}^2}{2\gamma_{{\rm lp}, \textbf{k}} } = \frac{\Omega_e^2}{2\gamma}\left( \frac{\omega_{\textbf{k}}}{\omega_{\textbf{k}, {\rm lp}}} \right)^2, \\\label{eq:Gamma_gamma_dominated_up}
    \Gamma_{\rm up, \textbf{k}}^{\gamma} &\approx& \frac{\Omega_{\rm up, \textbf{k}}^2}{2\gamma_{{\rm up}, \textbf{k}} } = \frac{\Omega_e^2}{2\gamma}\left( \frac{\omega_{\textbf{k}}}{\omega_{\textbf{k}, {\rm up}}} \right)^2.
\end{eqnarray}
\iac{As the photonic weight enters in both the decay rates and in the coupling to the emitter, the information about the USC mixing angle has almost completely disappeared and one is left with a relatively weak dependence on the polariton frequency. In the strong but not ultra-strong coupling regime, this gives equal decay rates (\ref{eq:Gamma_gamma_dominated_lp}-\ref{eq:Gamma_gamma_dominated_up}) for respectively the lower and upper polariton resonances; in the ultra-strong coupling regime, the lower polariton has a stronger decay rate.} 

In the second case, more relevant for ISB systems \cite{DeBernardis_PhysRevB.106.224206}, we have
\begin{eqnarray}
    \gamma_{\text{lp}, \textbf{k}} &=& \kappa_{\rm d} \sin^2 \theta_{\textbf{k}}, \\
    \gamma_{\text{up}, \textbf{k}} &=& \kappa_{\rm d} \cos^2 \theta_{\textbf{k}}.
\end{eqnarray}
Here $\kappa_{\rm d}$ is the bare decay rate of the uncoupled dresser. 
As for the previous case, the dresser-dominated decay rate is then given by
\begin{eqnarray}\label{eq:Gamma_kappa_dominated_lp}
    \Gamma_{\rm lp, \textbf{k}}^{\kappa_{\rm d}} \approx \frac{\Omega_{\rm lp, \textbf{k}}^2}{2\gamma_{{\rm lp}, \textbf{k}} } &=& \frac{\Omega_e^2}{2\kappa_{\rm d}}\left( \frac{\omega_{\textbf{k}}}{\omega_{\textbf{k}, {\rm lp}}} \right)^2 \cot^2 \theta_{\textbf{k}} , \\
\label{eq:Gamma_kappa_dominated_up}
    \Gamma_{\rm up, \textbf{k}}^{\kappa_{\rm d}} \approx \frac{\Omega_{\rm up, \textbf{k}}^2}{2\gamma_{{\rm up}, \textbf{k}} } &=& \frac{\Omega_e^2}{2\kappa_{\rm d}}\left( \frac{\omega_{\textbf{k}}}{\omega_{\textbf{k}, {\rm up}}} \right)^2 \tan^2 \theta_{\textbf{k}} .
\end{eqnarray}
\iac{In the strong coupling regime, the difference between the two rates is determined by the trigonometric functions in the mixing angle. In the ultra-strong coupling regime, there is the additional frequency-dependent factor favouring the decay in the lower polariton resonance case.}

\iac{All together, these formulas show how, in analogy} with the corresponding modification of the Rabi splitting in the strong coupling regime, the effect of the USC vacuum state in the weak emitter-polariton coupling regime will be to enhance (suppress) the spontaneous emission into the lower (upper) polariton by an amount related to the modified quantum fluctuations of the electric displacement field as shown in Eq. \eqref{eq:vacuum}.

In physical terms, Eqs. (\ref{eq:Gamma_gamma_dominated_lp}-\ref{eq:Gamma_gamma_dominated_up}) and  (\ref{eq:Gamma_kappa_dominated_lp}-\ref{eq:Gamma_kappa_dominated_up}) can be understood as novel forms of the Purcell effect~\cite{purcell_PhysRev.69.37, Englund_PhysRevLett.95.013904}, where the modification acts on the overall intensity of the field fluctuations in the USC vacuum rather than merely their frequency redistribution. 
Interestingly, our result aligns with the theory of spontaneous emission in dielectric media and has the key advantage of a clear disentanglement of local field effects~~\cite{Lewenstein:PRA1991,Mukamel:PRA1989,Scheel:PRA1999,Barnett_PhysRevLett.68.3698, Berman_PhysRevLett.92.053601, Lagendijk_PhysRevLett.81.1381}.
Beside the mere academic interest, this predicted effect can have a strong technological impact too, since spontaneous emission and Purcell factors give for instance the basis for the theory of laser's linewidths \cite{Milonni_PhysRevA.44.1969}, or being also relevant for quantum non-linear optics \cite{Purcell_SPDC1_10.1063/1.5044539, Purcell_SPDC2_Davoyan:18}.

%, Poddubny_PhysRevB.87.035136}.
\subsection{Non-linear optics}
Even though our discussion is carried out only in \iac{the linear-optics regime of bosonic ISB polaritons,} changing the quantum vacuum amplitude of the electric displacement field has strong consequences also in non-linear optics \iac{effects at strong illumination levels where polariton-polariton interactions become important~\cite{WallsMilburn,carusotto_RevModPhys.85.299}.}

Indeed, the standard theory of non-linear optics~\cite{hong_mandel_PhysRevA.31.2409} starts by considering the interaction Hamiltonian
\begin{equation}\label{eq:ham_emitter_D_general}
H_{e,\,I} = - \int d^3 r~  \frac{\textbf{D}(\textbf{r})\cdot \textbf{P}_{e}(\textbf{r})}{\epsilon_0} + \int d^3 r ~ \frac{\textbf{P}^2_{e}(\textbf{r})}{2 \epsilon_0}.
\end{equation}
of the electromagnetic field coupled to the non-linear medium, which, by continuity, here is also labelled as the emitter.
Introducing the non-linear polarizability tensor $\chi^{(n)}_{i_1\, i_2\, \ldots i_n}$ of order $n$, we can rewrite the the emitter polarization density as 
\begin{equation}
    \textbf{P}_{e}(\textbf{r}) \sim \int d^3 r~ \chi^{(n)}_{i_1\, i_2\, \ldots i_n}  D_{i_1}(\textbf{r})\, D_{i_2}(\textbf{r}) \cdots D_{i_n}(\textbf{r})\,.
\end{equation} 
Combining it with the general expression of Eq. \eqref{eq:displacement_polariton_basis1}
%\eqref{eq:electric_displacement_pol_basis_zp_coefficient} (or the specific expression for ISB polaritons in Eq. \eqref{eq:displacement_polariton_basis1}) 
it is then clear that also \iac{nonlinear optical processes involving quantum vacuum fluctuations can be either boosted or suppressed} by the modified zero-point electric displacement amplitude.

Taking as the simplest example the \iac{spontaneous parametric downconversion effects mediated by a $\chi^{(2)}$} non-linearity, in the ISB polariton case we obtain the three-wave mixing Hamiltonian
\begin{equation}
    H_{3\rm wm} \sim \!\!\!\sum_{ \substack{\textbf{k}_1, \textbf{k}_2, \textbf{k}_3 \\\ell_1, \ell_2, \ell_3= {\rm lp, up}} } \int d \omega \,\tilde{\chi}^{(2)}_{\rm TM_0}\, p_{\ell_1,\, \textbf{k}_1}^{\dag} p_{\ell_2,\, \textbf{k}_2}^{\dag} p_{\ell_3,\, \textbf{k}_3} + {\rm h.c.}
\end{equation}
where the \iac{effective nonlinearity scales as}
\begin{equation}
    \tilde{\chi}^{(2)}_{\rm TM_0} = \chi^{(2)}_{\rm TM_0} \left(\frac{\epsilon_0 \hbar \omega_e}{2SL_c \Omega_e^2}\right)^3 \Omega_{\ell_1, \textbf{k}_1} \Omega_{\ell_2, \textbf{k}_2} \Omega_{\ell_3, \textbf{k}_3},
\end{equation}
\iac{where momentum and energy conservations impose that} $\textbf{k}_1 + \textbf{k}_2 = \textbf{k}_3$ and $\omega_{\ell_1, \textbf{k}_1} + \omega_{\ell_2, \textbf{k}_2} = \omega_{\ell_3, \textbf{k}_3}$.

\iac{This formula shows that the strength of the nonlinear processes is also modified according to Fig. \ref{fig:2}(a). Similarly to the emitter-emitter interaction and the spontaneous emission rate, also the parametric downconversion process appears to be reinforced for the lower polariton branch.}

%\cite{hong_mandel_PhysRevA.31.2409, milonni1994quantum, milonni_atoms7010027}

\section{Relation with classical theory of dielectrics}
\label{sec:relation_classical_dielectric}

Our previous derivations were carried out within a quantum language, so the modifications of the vacuum-field Rabi splitting, the effective emitter interactions and of the spontaneous emission rate were naturally related to the ones of the zero-point fluctuations \eqref{eq:vacuum}.
In connection to the intense debate that took place on the physical origin of spontaneous emission in terms of quantum fluctuation and/or radiative reaction effects~\cite{Senitzky_PhysRevLett.31.955,milonni1973interpretation,dalibard:jpa-00209544}, it is interesting to assess whether our predictions can be equivalently expressed in classical terms of radiative reaction by the dressed electromagnetic field in the USC device. 

As a first step in this direction, it is straightforward to notice that our theory can be directly reformulated in terms of motion equations for the cavity $a_{\textbf{k}}$ and the emitter $b_{\textbf{k}}$ and dresser $d_{\textbf{k}}$ polarization fields stemming from the Hamiltonian \eqref{eq:ham_plasma_emitter} and re-interpreted as classical variables. 
%After some lengthy algebra reported in Secs.\ref{app:classical_trans_spectra}-\ref{app:dipolar_cQED_classical} of the SM, the eigenmodes of 
This set of equations can be summarized in a dispersion relation (that can be equivalently derived directly from Maxwell equations, see App. \ref{app:dipolar_cQED_classical})
%\ref{app:classical_trans_spectra}-\ref{app:dipolar_cQED_classical} of the SM) 
of the form $\omega^2\,\epsilon_r(\omega)=c^2 k^2$, with the total effective dielectric permittivity for the system as a whole
%$\epsilon_r(\omega)^{-1} = 1-\frac{\Omega_d^2}{\bar{\omega}_d^2-\omega^2}-\frac{\Omega_e^2}{\bar{\omega}_e^2-\omega^2} $,
\begin{equation}
    \epsilon_r^{\rm tot}(\omega) = \frac{1}{1-\frac{\Omega_d^2}{\bar{\omega}_d^2-\omega^2 - i\kappa_{\rm d}\omega}-\frac{\Omega_e^2}{\bar{\omega}_e^2-\omega^2- i\kappa_{e}\omega}}\, ,
\end{equation}
that is an analog of the \emph{Clausius-Mossotti} formula in a multiple slab geometry \cite{Hopfield_PhysRev.112.1555,Bottcher_1974} (here $\kappa_{\rm d}, \kappa_{e}$ represent the dresser and emitter losses). 
%In Sec. \ref{app:dipolar_cQED_classical} of the SM is shown that this expression can be equivalently derived from classical Maxwell equations of dielectrics.
Interestingly, the deviation of this expression from a naive simple sum of the individual slabs' permittivities, 
$\epsilon_r^{\rm tot}(\omega ) \approx 1 + \Omega_{\rm d}^2/({\omega}_{\rm d}^2 - \omega^2 - i\kappa_{\rm d}\omega) + \Omega_e^2/({\omega}_e^2 - \omega^2 - i\kappa_{e}\omega)$ 
%$\epsilon_r^{\rm tot}(\omega ) \approx 1 + \Omega_{\rm d}^2/(\bar{\omega}_{\rm d}^2 - \omega^2 - i\kappa_{\rm d}\omega) + \Omega_e^2/(\bar{\omega}_e^2 - \omega^2 - i\kappa_{e}\omega)$ 
provides a signature of the key role played by the electrostatic interaction between the emitter and the dresser noticed in~\cite{DeBernardis_cavityQED_nonperturbative_PhysRevA.97.043820,mazza_amelio_PhysRevB.104.235120, rocio_2022observe,pantazopoulos2023electrostatic}, whose strength is indeed proportional to the product of the electron densities $n_{d}n_e\sim \Omega^2_d\Omega_e^2$. 
%Once again, $\bar{\omega}^2_{d,e}={\omega}^2_{d,e}+\Omega_{d,e}^2$ are the depolarization-shifted values of the dresser and emitter frequencies~\cite{zaluzny1982inter,ando_electronic_1982,cominotti2023theory}.

\iac{Within this framework, formal integration of the motion equations for the cavity $a_{\textbf{k}}$ and dresser $d_{\textbf{k}}$ fields (or, equivalently, the cavity-dresser polaritons) as a function of the emitter field $b_{\textbf{k}}$ leads to effective terms in the equation of motion for $b_{\textbf{k}}$ that can are straightforwardly interpreted as radiative reaction.
In the strong emitter-polariton coupling regime, this formulation in terms of classical equations of motion generalizes the standard radiative reaction equations~\cite{Senitzky_PhysRevLett.31.955,dalibard:jpa-00209544} to the case of a resonantly peaked density of states of the electromagnetic modes (see App. \ref{app:spont_em})}: \ddb{the strong frequency-dependence of the radiative reaction results in the splitting of the emitter's frequency into several polariton modes.}

\iac{In the weak emitter-polariton coupling regime, for an emitter close to resonance to the lower (upper) polariton mode,  this same procedure leads to a radiative damping term in the usual form (see App.\ref{app:dipolar_cQED_classical})
%\ref{app:dipolar_cQED_classical} of the SM)
%\begin{equation}
%\dot{b}_{\textbf{k}}= \ldots -\frac{\Omega_{\rm lp (up), \textbf{k}}^2}{2\gamma_{\rm lp (up)}}\,b_{\textbf{k}}\,.
%\end{equation}
\begin{equation}
    \dot{b}_{\textbf{k}}= -i\omega_eb_{\textbf{k}} -\frac{\Omega_{\rm lp (up), \textbf{k}}^2}{2\gamma_{\rm lp (up), \textbf{k}}}\,b_{\textbf{k}}\,,
\end{equation}
in agreement with the formulas (\ref{eq:Gamma_gamma_dominated_lp}-\ref{eq:Gamma_gamma_dominated_up}) and (\ref{eq:Gamma_kappa_dominated_lp}-\ref{eq:Gamma_kappa_dominated_up}).}

\iac{In general, this} close correspondence between the radiative reaction and the quantum fluctuations pictures is a direct manifestation of the fluctuation-dissipation theorem relating the quantum fluctuations of the dressed polariton field in its vacuum state, that is the correlation function \eqref{eq:Greenij}, to the susceptibility of vacuum in response to the emitter's polarization~\cite{Barnett_PhysRevLett.68.3698,dalibard:jpa-00209544}. 

Altogether, these arguments show how the entire vacuum phenomenology described so far can be equivalently re-interpreted in terms of electrodynamics in dense dielectric media. 
While the complete equivalence between this approach and the polaritonic approach is now clear, it is worth noting that the polaritonic approach allows for a simpler and more systematic implementation of suitable approximations for the system. This constitutes an important tool to unpack all the relevant contributions to the otherwise challenging solution of the dyadic wave equation in presence of materials \cite{Chang_RevModPhys.90.031002}.

\section{Conclusions and outlook}
\label{sec:conclusion}
In this work, we have theoretically investigated how basic light-matter interaction processes are modified in the distorted vacuum state of a semiconductor-based cavity QED system in the ultra-strong coupling (USC) regime.  
In particular, we have focused our attention on quantities of direct experimental access %in cavity transmission spectra, 
such as the vacuum-field Rabi splitting of an additional emitter strongly coupled to USC polaritons, and its spontaneous emission in the weak emitter-polariton coupling limit. In both cases,
a signature of the distorted vacuum state of the USC regime is visible as a marked asymmetry between the polariton branches.

Even though our discussion is focused on a specific material system of major experimental interest, the predicted effects generally apply to any optical system in the USC regime. As such, our conclusions are of interest for a broad community of researchers, from circuit-QED devices, to semiconductor optoelectronics and terahertz optics and validate the picture that engineering the QED vacuum is indeed a powerful tool to control optical processes and, on the longer run, possibly also manipulate the \iac{electronic} properties of materials \cite{ebbesen_ciuti_review, Jacqueline_mohammad_review,andolina_PhysRevB.102.125137,andolina2022deep}. 

From a conceptual standpoint, we have shown that our predictions can be equivalently understood in terms of classical radiative reaction in dielectric materials or in terms of quantum fluctuations in the distorted USC vacuum state, the two pictures being connected by the fluctuation-dissipation theorem.
On one hand, the connection to classical radiative reaction 
provides a new point of view on vacuum effects as a tool to control materials, a concept that is attracting a growing interest, but is typically investigated within a quantum language~\cite{Ashida_PhysRevX.10.041027, ashida_VdW_PhysRevLett.130.216901, geva_PhysRevB.107.045425, curtis2023local, ebbesen_ciuti_review, Jacqueline_mohammad_review, svendsen2023theory}.
On the other hand, the quantum picture provides a transparent framework to go beyond a Fermi golden rule analysis of spontaneous emission in dielectric materials and describe the interplay of ultra-strong coupling with more sophisticated light-matter interaction phenomena such as parametric downconversion, which directly involve the amplification of quantum fluctuations. 
%On the other hand,  

\acknowledgements{ We are grateful to Raffaele Colombelli, Francesco Pisani, Yanko Todorov, Marco Schir\`o, Simone De Liberato, Peter Rabl and Giuseppe La Rocca for useful discussions. We acknowledge funding from the European Research Council (ERC) under the European Union's Horizon 2020 research and innovation
programme (Grant agreement No. 101002955 -- CONQUER), from the Provincia Autonoma di Trento, the Q@TN initiative, and the PNRR MUR project PE0000023-NQSTI. D.D.B. acknowledges funding from the European Union - NextGeneration EU, "Integrated infrastructure initiative in Photonic and Quantum Sciences" - I-PHOQS [IR0000016, ID D2B8D520, CUP B53C22001750006]. }

\appendix

\section{Intersubband polaritons in the dipole representation}
\label{app:cQED_ham_derivation}

We review here the details of the derivation of the intersubband (ISB) polariton Hamiltonian valid in the ultrastrong coupling (USC) regime, based on Ref. \cite{Todorov_2012_PhysRevB.85.045304}.
Initially we focus only on a single quantum well (QW) and in the next section we generalize it to a multi-well configuration.

The polarization density can be written explicitly in terms of raising/lowering operators of the respective intersubband transition \cite{Todorov_2012_PhysRevB.85.045304}
\begin{equation}
\textbf{P}(\textbf{r}) = \frac{e}{S}\sum_{\mu >\nu, \textbf{k}} \zeta_{\mu \nu}(z) e^{i \textbf{k}\cdot \textbf{r}_{\parallel}} \left( B_{\mu \nu \textbf{k}} + B_{\mu \nu -\textbf{k}}^{\dag} \right) \textbf{u}_z,
\end{equation}
Here $e$ is the electron charge, $S$ is the surface area of the slab (assumed to be equal to the cavity) and $\textbf{u}_z$ is a unit vector along the $z$-axis. The in-plane position is $\textbf{r}_{\parallel} = (x,y,0)$ and $\textbf{k} = (k_x, k_y, 0)$ is the in-plane wavector.
The dipole density is given by 
\begin{equation}
    \zeta_{\mu \nu}(z) = z \psi_{\mu}^*(z)\psi_{\nu}(z).
\end{equation}
Here $\psi_{\mu}(z)$ is the wavefunction of the $\mu$-th level of the QW \cite{Todorov_2012_PhysRevB.85.045304, Todorov_2015_dipolar_QED_2D_egas_PhysRevB.91.125409, Todorov_dipolar_QED_3D_egs_PhysRevB.89.075115}, defined in the interval $[z_{\rm qw}-L_{\rm qw}/2, z_{\rm qw}+L_{\rm qw}/2 ]$, where $L_{\rm qw}$ is the size of the quantum well along the $z$-axis and $z_{\rm qw}$ is its central position.
The ISB transition operators are given by
\begin{equation}
    B_{\mu \nu \textbf{k}} = \sum_{\textbf{q}}\tilde{\Psi}^{\dag}_{\nu, \textbf{q}-\textbf{k}}\tilde{\Psi}_{\mu,\textbf{q}},
\end{equation}
where $\tilde{\Psi}_{\mu, \textbf{k}}$ is the Fermionic operator that annihilates an electron with in-plane momentum $\textbf{k}$ in the $\mu$-th subband.
It is convenient to rewrite this operator as
\begin{equation}
B_{\mu \nu \textbf{k}} = \sqrt{N_{\mu \nu}} b_{\mu \nu \textbf{k}}.
\end{equation}
where $N_{\mu \nu} = \braket{\hat{N}_{\nu}} - \braket{\hat{N}_{\mu}}$ is the electron population imbalance between two subbands. 
Considering only low energy transitions which take place just from the ground state (Fermi sea) to an upper band and assuming $N_{\mu \nu} \gg 1$ (heavy doping), we have that 
\begin{equation}\label{eq:bosonic_comm_ISB}
    [b_{\mu \nu \textbf{k}}, b_{\mu ' \nu ' \textbf{k}\, '}^{\dag}] \simeq \delta_{\mu \mu '} \delta_{\nu \nu '} \delta_{ \textbf{k}\, \textbf{k}\, '}
\end{equation}
and the transition operators behave as Bosonic creation/annihilation operators for the collective excitation. 
Considering only transitions from the lowest subband, in the low excitation limit, we have that $N_{\mu \nu=0} \approx N$, where $N$ is the total number of electrons. From now on we always consider only transitions from the lowest subband, suppressing the index $\nu$.
The QW Hamiltonian is then just given by a collection of harmonic oscillators
\begin{equation}
    H_{\rm qw} = \hbar \omega_{\rm qw} \sum_{\textbf{k}} b_{\textbf k}^{\dag} b_{\textbf k}.
\end{equation}

The light-matter Hamiltonian for the QW coupled to the cavity field can be derived by considering the total energy of the system, which is given by the matter's energy plus the electromagnetic energy \cite{jackson_classical_1999}
\begin{equation}
\begin{split}
    H &= H_{\rm qw} + H_{\rm em}
    \\
    & = H_{\rm qw} + \int d^3 r \left[ \frac{\epsilon_0\textbf E^2(\textbf r) }{2} + \frac{\epsilon_0 c^2}{2}\textbf B^2(\textbf{r})  \right].
\end{split}
\end{equation}
The coupling between matter and the electromagnetic field is provided by the fact that the charged matter generates and changes the electric and magnetic field. 
In the so-called dipole gauge this is realized by the following minimal coupling substitution in the cavity electric field energy density \cite{jackson_classical_1999}
\begin{equation}
\epsilon_0 \textbf{E}(\textbf{r}) = \textbf{D}(\textbf{r}) - \textbf{P}(\textbf{r}).
\end{equation}
Notice that here we take only the displacement field to be transverse \cite{Cohen-Tannoudji:113864}, such that $\vec{\nabla}\cdot \textbf{D}(\textbf{r}) = 0$.
In this way we correctly recover Maxwell equations with a polarizable medium, where $\vec{\nabla}\cdot \textbf{E}(\textbf{r}) = - \vec{\nabla}\cdot\textbf{P}(\textbf{r})/\epsilon_0 = \rho_{b}(\textbf{r})/\epsilon_0$, where $\rho_{b}(\textbf{r})$ is the so-called bound charge density \cite{jackson_classical_1999}.

The total light-matter Hamiltonian is then given by \cite{Cohen-Tannoudji:113864}
%{\color{blue}\bf DDB:check dipoleGauge for localized dipole still works and all self energies and Coulomb do the same as for localized dipoles...the slabs do actually produce electric field and image charges on the cavity plates (see Eq. A.50 of my thesis), but still they do not talk directly through this electric field, which is cancelled by the residual Coulomb in the dipole gauge (and here must be careful in keeping trace of P or P transverse, cause the term P1P2 is probably meant to be transfer and then exctly cancelled by the Coulomb one)}
\begin{equation}\label{eq:Ham_LM_tot_app}
\begin{split}
H = H_{\rm qw} + \int d^3 r \left[ \frac{\left( \textbf{D}(\textbf{r}) - \textbf{P}(\textbf{r}) \right)^2}{2 \epsilon_0} - \frac{\epsilon_0 c^2}{2}\textbf{A}(\textbf{r})\cdot \nabla^2 \textbf{A}(\textbf{r})  \right].
\end{split}
\end{equation}

The electric displacement in the cavity made of two perfect parallel mirrors can be written as \cite{jackson_classical_1999, rocio_2022observe}
\begin{equation}
\begin{split}
    \textbf{D}(\textbf{r}) = & i \sum_{n,\textbf{k},\lambda} \sqrt{\frac{\epsilon_0 \hbar\omega_{n,\textbf{k}}}{2 S L_c}} \\
    &\times \Bigg[ \textbf{w}_{n,\textbf{k},\lambda}(\textbf{r}_{\parallel}, z) a_{n,\textbf{k}} -  \left(\textbf{w}_{n,\textbf{k},\lambda}(\textbf{r}_{\parallel}, z) \right)^* a_{n, -\textbf{k}}^{\dag}\Bigg],
\end{split}
\end{equation}
where $S=L_xL_y$ is the surface area of the rectangular cavity and $L_c$ is the cavity height.
The adimensional cavity mode functions are solutions of the Poisson equation and can be written as \cite{rocio_2022observe}
\begin{equation}
    \textbf{w}_{n,\textbf{k},\lambda}(\textbf{r}_{\parallel}, z) = e^{i \textbf{k}\cdot \textbf{r}_{\parallel}}\sqrt{\frac{2}{\left(1+\delta_{n,0}\right)}} 
    \begin{pmatrix}
        i\varepsilon^{(x)}_{n,\textbf{k},\lambda}\sin (k_n z)\\
        i\varepsilon^{(y)}_{n,\textbf{k},\lambda}\sin (k_n z)\\
        \varepsilon^{(z)}_{n,\textbf{k},\lambda}\cos (k_n z)\\
    \end{pmatrix}
    ,
\end{equation}
where $k_n=\pi/L_c n$, with $n=0,1,2\ldots$ is the wavevector along the cavity axis, $\vec \varepsilon_{n,\textbf{k},\lambda}$ is the polarization vector and $\lambda=1,2$ is the polarization index.
Here metallic boundary conditions are assumed. 

Considering only ISB transitions from the lowest subband, $\nu = 0$, the interaction light-matter Hamiltonian for a single QW placed at $z=z_{\rm qw}$ is
\begin{equation}\label{eq:Ham_int_total}
\begin{split}
&H_I   = \int d^3 r \frac{1}{\epsilon_0}\left[ - \textbf{D}\cdot \textbf{P} + \frac{P^2}{2} \right] 
\\
& = -i\frac{\hbar \omega_P }{2} \sum_{n,\textbf{k},\lambda, \mu}\varepsilon^{(z)}_{n,\textbf{k},\lambda}\sqrt{f^n_{\mu 0}\frac{\omega_{n,\textbf{k}}}{\omega_{\mu 0}} } (a_{n, \textbf{k}} -  a_{n, -\textbf{k}}^{\dag}) \left( b_{\mu -\textbf{k}} + b_{\mu \textbf{k}}^{\dag} \right)
\\
& + \frac{\hbar \omega_P^2 }{4}\sum_{\textbf{k}}  \sum_{\mu, \mu'} \frac{I_{\mu \mu '}}{\sqrt{\omega_{\mu 0}\omega_{\mu' 0}}}\left(b_{\mu \textbf{k}} + b_{\mu -\textbf{k}}^{\dag}\right) \left(b_{\mu' \textbf{k}} + b_{\mu' -\textbf{k}}^{\dag}\right) 
\end{split}
\end{equation}
Here the main coupling parameter is \ddb{the full-filled cavity} \emph{plasma frequency}
\begin{equation}\label{eq:plasma_freq}
\omega_P^2 = \frac{N e^2}{\epsilon_0 m SL_c},
\end{equation} 
and $m$ is the electron effective mass. \ddb{Notice that the usual plasma frequency is defined only including the effective length of the QW $L_{\rm qw, eff}$, as done in Ref. \cite{Todorov_2012_PhysRevB.85.045304}. As a consequence in Eq. \eqref{eq:Ham_int_total} one has to include a cavity filling $\sim L_{\rm qw, eff}/L_c$. In order to minimize the definitions of new parameters, we simply define the plasma frequency at full filling.}
The generalized oscillator strength is defined as
\begin{equation}
\begin{split}
    f^n_{\mu 0} = \frac{2m\omega_{\mu 0} }{\hbar}\frac{2}{1+\delta_{n0}} \left[ \int_{-\infty}^{\infty}dz \zeta_{\mu 0}(z)\cos\left( k_n z \right) \right]^2.
\end{split}
\end{equation}
For $n=0$ it coincides with the usual oscillator strength of the $\mu$-th dipole transition $f_{\mu 0}= 2m \omega_{\mu 0} z_{\mu 0}^2/\hbar$, where 
\begin{equation}
    z_{\mu 0} = \braket{\mu | z | 0} = \int_{-\infty}^{\infty}dz \zeta_{\mu 0}(z)
\end{equation}
is the ISB dipole matrix element. At higher $n>0$, it also contains all the multipoles matrix elements, as it has to be by interacting with a non-homogeneous electric field. The higher the $n$-th mode is the more multipolar will be the ISB excitation.

In the last term of Eq. \eqref{eq:Ham_int_total} we introduced the $P^2$-interaction tensor strength
\begin{equation}
\begin{split}
    &I_{\mu \mu '} = \frac{2m L_c}{\hbar^2}\omega_{\mu 0}\omega_{\mu' 0} \int_{-\infty}^{\infty}dz \zeta_{\mu 0}(z)\zeta_{\mu' 0}(z)
    \\
    & = \frac{2m L_c}{\hbar^2}\omega_{\mu 0}\omega_{\mu' 0} \int_{-\infty}^{\infty}dz dz' \delta(z-z')\zeta_{\mu 0}(z)\zeta_{\mu' 0}(z')
    \\
    & = \frac{2m}{\hbar^2}\omega_{\mu 0}\omega_{\mu' 0} \frac{L_c}{2\pi}\int_{-\infty}^{\infty} dq_z \int_{-\infty}^{\infty}dz \zeta_{\mu 0}(z) \chi_{q_z}^*(z)   \times
    \\
    &\int_{-\infty}^{\infty}dz' \zeta_{\mu ' 0}(z') \chi_{q_z}(z'), 
\end{split}
\end{equation}
where in the last equality we introduced a complete basis such that $\int_{-\infty}^{\infty} dq_z/(2\pi)\chi_{q_z}^*(z)\chi_{q_z}(z') = \delta (z-z')$.

%In the last term of Eq. \eqref{eq:Ham_int_total} we introduced the $P^2$-interaction tensor strength
%\begin{equation}
%\begin{split}
%    &I_{\mu \mu '} = \frac{2m L_c}{\hbar^2}\omega_{\mu 0}\omega_{\mu' 0} \int_{-\infty}^{\infty}dz \zeta_{\mu 0}(z)\zeta_{\mu' 0}(z)
%    \\
%    & = \frac{2m L_c}{\hbar^2}\omega_{\mu 0}\omega_{\mu' 0} \int_{-\infty}^{\infty}dz dz' \delta(z-z')\zeta_{\mu 0}(z)\zeta_{\mu' 0}(z')
%    \\
%    & = \frac{2m L_c}{\hbar^2}\omega_{\mu 0}\omega_{\mu' 0} \sum_{n=0} \int_{-\infty}^{\infty}dz \zeta_{\mu 0}(z) \chi_n(z)   \times
%    \\
%    &\int_{-\infty}^{\infty}dz \zeta_{\mu ' 0}(z) \chi_n(z), 
%\end{split}
%\end{equation}
%where in the last equality we introduced a complete basis such that $\sum_{n=0}\chi_n(z)\chi_n(z') = \delta (z-z')$.

Truncating the ISB transitions to $\mu=1$ only, we suppress also the index $\mu$ having $b_{\mu, \textbf{k}}\mapsto b_{\textbf{k}}$ and defining $\omega_{\rm qw} = \omega_{10}$ and $f^n_{\rm qw} = f^n_{1 0}$. The interaction Hamiltonian becomes
\begin{equation}
    \begin{split}
&H_I \approx -i\frac{\hbar \omega_P }{2} \sum_{n, \textbf{k},\lambda}\varepsilon^{(z)}_{n,\textbf{k},\lambda}\sqrt{f^n_{\rm qw}\frac{\omega_{n,\textbf{k}}}{\omega_{\rm qw}} } (a_{n, \textbf{k}} -  a_{n, -\textbf{k}}^{\dag}) \left( b_{ -\textbf{k}} + b_{ \textbf{k}}^{\dag} \right) 
\\
&+ \frac{\hbar \omega_P^2 }{4\omega_{\rm qw}}\sum_{\textbf{k}} \frac{L_c}{2\pi}\int_{-\infty}^{\infty} dq_z \tilde{f}_{\rm qw}(q_z)\left(b_{ \textbf{k}} + b_{ -\textbf{k}}^{\dag}\right) \left(b_{ \textbf{k}} + b_{ -\textbf{k}}^{\dag}\right) 
\end{split}
\end{equation}
where \ddb{we have introduced the continuous version of the generalized oscillator strength 
\begin{equation}
    \tilde{f}_{\rm qw}(q_z) = \frac{2m\omega_{\rm qw}}{\hbar}\left|\int_{-\infty}^{\infty}dz \zeta_{\mu 0}(z) \chi_{q_z}(z)\right|^2
\end{equation}
}
%we have used $\chi_n(z) = \sqrt{2/(L_c(1+\delta_{n,0}))}\cos (k_n z)$ \cite{Todorov_2012_PhysRevB.85.045304}.
\ddb{
Notice that in this Hamiltonian we have still included both the longitudinal and transverse contributions to the electric interactions due to the $\mathbf{P}^2$-term present in the original Hamiltonian in Eq. \eqref{eq:Ham_LM_tot_app}.
In order to single out the transverse part (which is the important one in the polaritonic physics) we rewrite the interaction Hamiltonian such that we separate the longitudinal and transverse contributions reading
\begin{equation}
    H_I = H_{\perp} + H_{\parallel}.
\end{equation}
Here
\begin{equation}\label{eq:Ham_finite_freq}
    \begin{split}
&H_{\perp} = -i\frac{\hbar \omega_P }{2} \sum_{n, \textbf{k},\lambda}\varepsilon^{(z)}_{n,\textbf{k},\lambda}\sqrt{f^n_{\rm qw}\frac{\omega_{n,\textbf{k}}}{\omega_{\rm qw}} } (a_{n, \textbf{k}} -  a_{n, -\textbf{k}}^{\dag}) \left( b_{ -\textbf{k}} + b_{ \textbf{k}}^{\dag} \right) 
\\
&+ \frac{\hbar \omega_P^2 }{4\omega_{\rm qw}}\sum_{\textbf{k}, n, \lambda} f_{\rm qw}^n |\varepsilon^{(z)}_{n,\textbf{k},\lambda}|^2  \left(b_{ \textbf{k}} + b_{ -\textbf{k}}^{\dag}\right) \left(b_{ \textbf{k}} + b_{ -\textbf{k}}^{\dag}\right), 
\end{split}
\end{equation}
and
\begin{equation}
    \begin{split}
&H_{\parallel} = \frac{\hbar \omega_P^2 }{4\omega_{\rm qw}}\sum_{\textbf{k}} \left( I_{11} - \sum_{\lambda, n}|\varepsilon^{(z)}_{n,\textbf{k},\lambda}|^2 f_{\rm qw}^n\right)  \left(b_{ \textbf{k}} + b_{ -\textbf{k}}^{\dag}\right) \left(b_{ \textbf{k}} + b_{ -\textbf{k}}^{\dag}\right).
\end{split}
\end{equation}
Using the polarization sum rule $\sum_{\lambda}|\varepsilon^{(z)}_{n,\textbf{k},\lambda}|^2 = 1 - k_n^2/k^2$ \cite{barton_1970,rocio_2022observe}, we obtain
\begin{equation}\label{eq:dep_shift_ES}
\begin{split}
    &H_{\parallel} = \frac{\hbar \omega_P^2 }{4\omega_{\rm qw}}\sum_{\textbf{k}}\left(b_{ \textbf{k}} + b_{ -\textbf{k}}^{\dag}\right) \left(b_{ \textbf{k}} + b_{ -\textbf{k}}^{\dag}\right)\times 
    \\
    & \times \left[ \sum_n f_{\rm qw}^n \frac{k_n^2}{k^2} + \left( \frac{L_c}{2\pi}\int_{-\infty}^{\infty} dq_z \tilde{f}_{\rm qw}(q_z) - \sum_n f_{\rm qw}^n  \right) \right].
\end{split}
\end{equation}
Notice that this contribution is nothing else than the longitudinal projection of the polarization density, restricted only to the $z$-direction, meaning the direct dipole-dipole Coulomb interaction within the slab. The first term of the second line, scaling with $k_n^2$ is the standard depolarization contribution due to the surface charges of the slab \cite{jackson_classical_1999}, while the second contribution, proportional to the difference between the integral and the sum, is the image charge contributions.
Moreover one can directly check using perturbation theory in the limit of $\omega_{\rm qw}\rightarrow 0$ or equivalently the adiabatic elimination of the cavity that all the contributions from the finite frequency Hamiltonian in Eq. \eqref{eq:Ham_finite_freq} cancels out \cite{DeBernardis_cavityQED_nonperturbative_PhysRevA.97.043820,rocio_2022observe}.
Since the contribution of Eq. \eqref{eq:dep_shift_ES} to the depolarization shift is irrelevant for our discussion, for the sake of simplicity, we will neglect it, considering it absorbed into the definition of $\omega_{\rm qw}$.
}

The ISB transition couple mostly with the so-called TM$_0$ mode \cite{Todorov_2015_dipolar_QED_2D_egas_PhysRevB.91.125409}, which corresponds to the $n=0$ mode, entirely polarized along the cavity axis ($z$-axis), with $\varepsilon^{(z)}_{0,\textbf{k},\lambda}=\delta_{\lambda,1}$. The light-matter Hamiltonian can be further simplified to
\begin{equation}\label{eq:ham_app_cavity-dresser}
    \begin{split}
&H \approx \hbar \omega_{\rm qw}\sum_{\textbf k} b_{\textbf k}^{\dag} b_{\textbf k} + \sum_{\textbf k}\hbar \omega_{\textbf k} a_{\textbf k}^{\dag} a_{\textbf k} 
\\
&-i\frac{\hbar \omega_P }{2} \sum_{\textbf{k}}\sqrt{ f_{\rm qw}\frac{\omega_{\textbf{k}}}{\omega_{\rm qw}} } (a_{\textbf{k}} -  a_{ -\textbf{k}}^{\dag}) \left( b_{ -\textbf{k}} + b_{ \textbf{k}}^{\dag} \right)
\\
&+ \frac{\hbar \omega_P^2 }{4\omega_{\rm qw}}\sum_{\textbf{k}} f_{\rm qw}\left(b_{ \textbf{k}} + b_{ -\textbf{k}}^{\dag}\right) \left(b_{ \textbf{k}} + b_{ -\textbf{k}}^{\dag}\right),
\end{split}
\end{equation}
where we have suppressed the $n$ index everywhere.
After all these passages we are left with an effective light-matter Hamiltonian describing the interaction with only the TM$_0$ cavity mode \cite{Todorov_2012_PhysRevB.85.045304}, which can be rewritten as
\begin{equation}
\begin{split}\label{eq:ham_eff_plasma}
&H_{{\rm TM}_0} = H_{\rm qw} 
\\
&+ L_c\int d^2 r \left[ \frac{\left( D(\textbf{r}_{\parallel}) - P(\textbf{r}_{\parallel}) \right)^2}{2 \epsilon_0} - \frac{\epsilon_0 c^2}{2}A(\textbf{r}_{\parallel}) \nabla^2 A(\textbf{r}_{\parallel})  \right].
\end{split}
\end{equation}
Here the displacement and the vector potential fields are scalar variables, meant to describe the $z$-component of the TM$_0$ modes, which are independent of the position along the $z$-axis (perpendicular to the cavity plane) and thus depends only from the in-plane position $\textbf r_{\parallel}$
\begin{equation}
    D(\textbf r_{\parallel} ) = i \sum_{\textbf{k}} \sqrt{\frac{\epsilon_0 \hbar\omega_{\textbf{k}}}{2SL_c}} e^{i \textbf{k}\cdot \textbf{r}_{\parallel}}(a_{\textbf{k}} -  a_{-\textbf{k}}^{\dag})
\end{equation}
\begin{equation}
   A(\textbf r_{\parallel} ) = \sum_{\textbf{k}} \sqrt{\frac{\epsilon_0 \hbar}{2SL_c\omega_{\textbf{k}}}} e^{i \textbf{k}\cdot \textbf{r}_{\parallel}}(a_{\textbf{k}} +  a_{-\textbf{k}}^{\dag}),
\end{equation}
with $[a_{\textbf{k}}, a_{\textbf{k}\,'}^{\dag}] = \delta_{\textbf{k},\textbf{k}\,'}$.
The effective polarization density coupled to the TM$_0$ modes is also a scalar quantity, and is given by
\begin{equation}
P(\textbf{r}_{\parallel}) = \frac{ez_{10}\sqrt{N}}{SL_c}\sum_{\textbf{k}}  e^{i \textbf{k}\cdot \textbf{r}_{\parallel}} \left( b_{\textbf{k}} + b_{-\textbf{k}}^{\dag} \right),
\end{equation}
It is worth noticing that nor the electromagnetic field densities nor the polarization densities of this TM$_0$-effective-description depends from the $z$-coordinate and so the overall physics does not depend from the position of the QW inside the cavity or its distance with respect to one or the other metallic plate.
This is in principle in contradiction with basics electromagnetism, where any charge configuration between metallic plates would generates image charges which will affect its behaviour as a function of the distance from the plate \cite{DeBernardis_cavityQED_nonperturbative_PhysRevA.97.043820}. This contradiction emerges as a consequence of the truncation to the TM$_0$ mode only, for which we have discarded all the information regarding the presence of image charges and eventual Coulombic corrections. However we argue that this corrections are most often negligible for our specific aims.

\section{Polariton Hamiltonian for two stacked wells}
\label{app:pol_ham_2_stacked_wells}
Since the polarization density of the matter inside the cavity is given by the sum over all its individual components, i.e. as a sum over all different quantum wells
\begin{equation}
\textbf{P}(\textbf{r}) =  \sum_{i=1}^{N_{\rm qw}}\textbf{P}_i(\textbf{r}),
\end{equation}
and each polarization densities do not overlap with the others $\int d^3r\textbf{P}_i(\textbf{r})\cdot \textbf{P}_j(\textbf{r}) = 0$ if $i\neq j$, we can easily generalize the Hamiltonian in Eq. \eqref{eq:ham_eff_plasma} to a multi-wells setup.
Notice that this condition of non-overlapping polarization densities is exactly the condition of localized dipoles used in the standard applications of the dipole picture \cite{Cohen-Tannoudji:113864}. As a consequence the direct Coulomb interaction between different QWs disappears in favour of a fully local mediated interaction through the dynamical cavity field.

In the simplest case of two QWs, the dresser and emitter, as in the main text, the resulting Hamiltonian is
\begin{equation}
\begin{split}
&H = H_{\rm d} + H_e + L_c\int d^2 r \left[ \frac{D^2}{2 \epsilon_0} - \frac{\epsilon_0 c^2}{2}A\nabla^2 A \right] 
\\
&+ L_c\int d^2 r \frac{1}{\epsilon_0}\left[ - D P_{\rm d} + \frac{P_{\rm d}^2}{2} \right] + L_c\int d^2 r \frac{1}{\epsilon_0}\left[ - D P_e + \frac{P_e^2}{2} \right].
\end{split}
\end{equation}
Since the two QWs have different electron number $N_{\rm d},N_e$, they also have different plasma frequencies.
These can be cast into the two Rabi frequencies
\begin{equation}
    \begin{split}
        &\Omega_{\rm d}^2 = f_{\rm d} \frac{N_{\rm d} e^2}{\epsilon_0 m SL_c},
        \\
        &\Omega_e^2 = f_e \frac{N_e e^2}{\epsilon_0 m SL_c}.
    \end{split}
\end{equation}
Notice that here we are using all the definitions of Sec. \ref{app:cQED_ham_derivation} replacing all the subscripts ${\rm qw}\mapsto {\rm d}, e$.

Introducing the creation-annihilation operators like in Sec. \ref{app:cQED_ham_derivation}, the total cavity QED Hamiltonian for the dresser and emitter QWs is thus given by
\begin{equation}\label{eq:ham_total_bosonic_rep}
\begin{split}
&H =  \omega_{\rm d}\sum_{\textbf k} d_{\textbf k}^{\dag}d_{\textbf k} + \omega_{e}\sum_{\textbf k} b_{\textbf k}^{\dag}b_{\textbf k} + \sum_{\textbf{k}} \omega_{\textbf{k}} a^{\dag}_{\textbf{k}}a_{\textbf{k}} 
\\
&-\frac{i\hbar \Omega_{\rm d}}{2} \sum_{\textbf{k}}\sqrt{\frac{\omega_{\textbf{k}}}{\omega_{\rm d}}} \left( a_{\textbf{k}} - a_{-\textbf{k}}^{\dag} \right)\left( d_{-\textbf{k}} + d_{\textbf{k}}^{\dag} \right) 
\\
&-\frac{i\hbar \Omega_{e}}{2} \sum_{\textbf{k}}\sqrt{ \frac{\omega_{\textbf{k}}}{\omega_{e}}} \left( a_{\textbf{k}} - a_{-\textbf{k}}^{\dag} \right)\left( b_{-\textbf{k}} + b_{\textbf{k}}^{\dag} \right) 
\\
& + \frac{\hbar \Omega_{{\rm d}}^2}{4\omega_{\rm d}} \sum_{\textbf{k}} \left( d_{-\textbf{k}} + d_{\textbf{k}}^{\dag} \right)\left( d_{-\textbf{k}} + d_{\textbf{k}}^{\dag} \right) 
\\
&+ \frac{\hbar \Omega_{{e}}^2}{4\omega_{e}} \sum_{\textbf{k}} \left( b_{-\textbf{k}} + b_{\textbf{k}}^{\dag} \right)\left( b_{-\textbf{k}} + b_{\textbf{k}}^{\dag} \right).
\end{split}
\end{equation}

Assuming that the emitter polarization density is always small, such that it never goes in the ultrastrong coupling regime we can neglect the $\sim P_e^2$-term of the emitter, which is the last term in Eq. \eqref{eq:ham_total_bosonic_rep}. In this way we recover the cavity-dresser-emitter Hamiltonian in the main text.

\section{Polaritons from canonical transformations}
\label{app:polariton_canonical_transform}
Here we consider the cavity-dresser Hamiltonian defined in Eq. \eqref{eq:ham_app_cavity-dresser}. Since the dresser Hamiltonian is quadratic, also the total cavity-dresser Hamiltonian is quadratic and can be thus diagonalized by a canonical transformation.
In this section we will suppress the vector notation for the in-plane wavevector $\textbf k \mapsto k$ in order to simplify the notation.

We then reintroduce the real canonical coordinates defining the quadrature operators
\begin{equation}\label{eq:def_canonical_quadratures}
\begin{split}
D_{k} & = -i\sqrt{\frac{\omega_k}{2}} \left( a_{k} - a^{\dag}_{-k} \right), \\
A_{k} & = \frac{1}{\sqrt{2\omega_k}}\left( a_{k} + a^{\dag}_{-k} \right), \\
\Pi_{k} & = -i\sqrt{\frac{\omega_{\rm d}}{2}} \left( d_k - d^{\dag}_{-k} \right),\\
X_{k} & =  \frac{1}{\sqrt{2\omega_{\rm d}}}\left( d_{k} + d^{\dag}_{-k} \right). \\
\end{split}
\end{equation}
The cavity-dresser Hamiltonian becomes
\begin{equation}
\begin{split}
    &H_{c-{\rm d}} = \sum_k \left[ \frac{\Pi_k^2}{2} + \frac{\omega_{\rm d}^2 + \Omega_{\rm d}^2}{2} X_k^2 \right] + 
    \\
    & +\sum_k \left[ \frac{D_k^2}{2} + \frac{\omega_{k}^2}{2} A_k^2 \right] + \sum_k \Omega_{\rm d} D_k X_k.
\end{split}
\end{equation}
We make a canonical transformation (just a re-labelling)
\begin{equation}
\begin{split}
\tilde{D}_k & = - \omega_k A_k, \\
\tilde{A}_k & = \frac{1}{\omega_k} D_k, \\
\end{split}
\end{equation}
We then have
\begin{equation}
\begin{split}
    &H_{c-{\rm d}} = \sum_k \left[ \frac{\Pi_k^2}{2} + \frac{\omega_{\rm d}^2 + \Omega_{\rm d}^2}{2} X_k^2 \right] +
    \\
    &+ \sum_k \left[ \frac{\tilde{D}_k^2}{2} + \frac{\omega_{k}^2}{2} \tilde{A}_k^2 \right] + \sum_k \Omega_{\rm d} \omega_k \tilde{A}_k X_k.
\end{split}
\end{equation}
The Hamiltonian in this form can be cast to a matrix, considering $H_{c-{\rm d}} = 1/2 v^T M v$, where $v^T = (X_k, \tilde{A}_k, \Pi_k, \tilde{D}_k) $, and
\begin{equation}
M =
\begin{pmatrix}
\omega_{\rm d}^2 + \Omega_{\rm d}^2 & \Omega_{\rm d} \omega_k & 0 & 0 \\
 \Omega_{\rm d} \omega_k & \omega_{k}^2& 0 & 0 \\
 0 & 0 & 1& 0 \\
 0 & 0 & 0 & 1 \\   
\end{pmatrix}
\end{equation}
Diagonalising $M$ is now equivalent to diagonalise the Hamiltonian. We can achieve this by considering the following unitary transformation
\begin{equation}
U =
\begin{pmatrix}
R & 0_{2\times 2}\\
0_{2\times 2} & R \\
\end{pmatrix}
\end{equation}
where $R$ is a $2\times 2$ rotation
\begin{equation}
R =
\begin{pmatrix}
 \cos \theta & \sin \theta \\
 -\sin \theta & \cos \theta \\
\end{pmatrix}
\end{equation}
In order to further simplify all the expressions we make implicit the dependence from the wavenumber $q$. Moreover we define
\begin{equation}
\begin{split}
&\Omega_X = \omega_{\rm d}^2 + \Omega_{\rm d}^2\\
& G  =  \Omega_{\rm d} \omega_k \\
& \Omega_A = \omega_{k}^2\\
\end{split}
\end{equation}
so
\begin{equation}
M = 
\begin{pmatrix}
\Omega_X & G & 0 & 0 \\
G & \Omega_A & 0 & 0 \\
 0 & 0 & 1& 0 \\
 0 & 0 & 0 & 1 \\   
\end{pmatrix}
\end{equation}
We then find
\begin{equation}\label{eq:sin_cos}
\begin{split}
\cos \theta & = \sqrt{ \frac{1}{2} \left( 1 + \frac{\Omega_X - \Omega_A}{\sqrt{\left( \Omega_X - \Omega_A \right)^2 + 4 G^2}} \right) } = \sqrt{\frac{\omega_{\rm up}^2-\omega_{k}^2}{\omega_{\rm up}^2 - \omega_{\rm lp}^2}} \\
\sin \theta & = \sqrt{ \frac{1}{2} \left( 1 - \frac{\Omega_X - \Omega_A}{\sqrt{\left( \Omega_X - \Omega_A \right)^2 + 4 G^2}} \right) } = \sqrt{\frac{\omega_{k}^2-\omega_{\rm lp}^2}{\omega_{\rm up}^2 - \omega_{\rm lp}^2}} \\
\end{split}
\end{equation}
with the diagonal matrix
\begin{equation}
M' =
\begin{pmatrix}
\omega_{\rm up}^2 & 0 & 0 & 0 \\
0 & \omega_{\rm lp}^2 & 0 & 0 \\
 0 & 0 & 1& 0 \\
 0 & 0 & 0 & 1 \\  
\end{pmatrix}
\end{equation}
where
\begin{equation}
\omega_{\rm up/lp}^2 = \frac{\Omega_X + \Omega_A}{2} \pm \frac{1}{2}\sqrt{\left( \Omega_X - \Omega_A \right)^2 + 4G^2}.
\end{equation}
Notice that this polariton spectrum is always real $\omega_{\rm up/lp}^2 > 0$, meaning that the TM$_0$ polariton Hamiltonian in Eq. \eqref{eq:ham_eff_plasma} is always stable, in accordance with the no-go theorems that forbids any cavity induced phase transitions \cite{Andolina_PhysRevB.100.121109, andolina_PhysRevB.102.125137, andolina_mercurio_2022}.
Working in the canonical representation has the convenience that, together with the eigenfrequencies, we have full access on the hybridization of the degrees of freedom due to the USC regime.
Differently from previous works \cite{nori_USC_review, Solano_USC_RevModPhys.91.025005}, we can condense the full knowledge of the four Hopfield coefficients \cite{Hopfield_PhysRev.112.1555, Ciuti_PhysRevB.72.115303}, in a single mixing angle $\theta_{ k}$.

\begin{figure}
    \centering
    \includegraphics[width=\columnwidth]{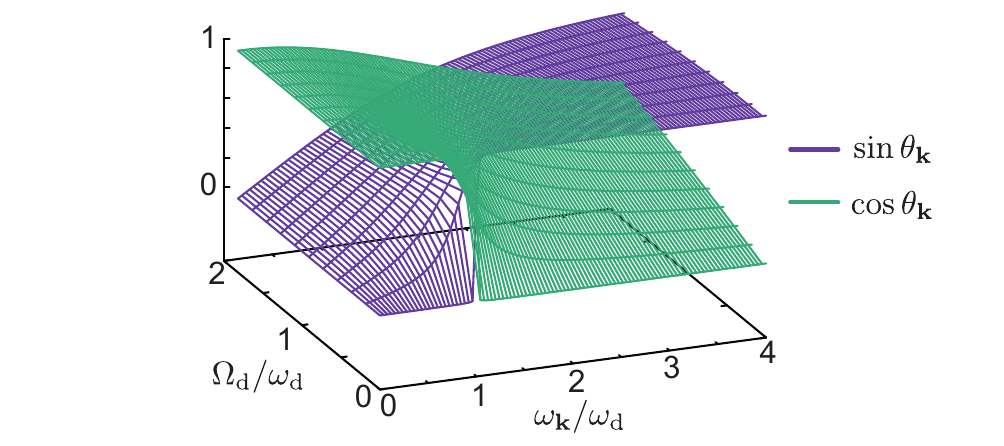}
    \caption{Canonical Hopfield coefficient $\sin\theta_{\textbf k},~\cos \theta_{\textbf k}$ as a function of the in-plane momentum and of the Rabi dressing $\Omega_{\rm d}$.}
    \label{fig:4}
\end{figure}

In Fig. \ref{fig:4} we show the behaviour of the  canonical Hopfield coefficients as a function of the in-plane wavenumber $k$ and the dresser Rabi frequency $\Omega_{\rm d}$. When the dresser is only weakly doped, and $\Omega_{\rm d}\ll \omega_{\rm d}$ we observe a very small hybridization ($\cos \sim 1$ and $\sin \sim 0$ or vice versa), which become substantial only when the TM$_0$ is resonant with the dresser frequency at $c k = \omega_{\rm d}$ (and $\cos \sim \sin \sim 1/2$). On contrary, when the dresser is in the USC regime $\Omega_{\rm d} \simeq \omega_{\rm d}$, the light-matter hybridization becomes important on a large range of wavenumbers.

The USC Hamiltonian in the new diagonal variables looks like two uncoupled harmonic oscillators, representing the upper/lower polaritons
\begin{equation}\label{eq:Ham_USC_diagonalised}
H_{c-{\rm d}} = \frac{\Lambda_{\rm up}^2}{2} + \frac{\omega_{\rm up}^2}{2}  \xi_{\rm up}^2 + \frac{\Lambda_{\rm lp}^2}{2} + \frac{\omega_{\rm lp}^2 }{2}\xi_{\rm lp}^2
\end{equation}
where
\begin{equation}\label{eq:canonical_relation_between_quadratures}
\begin{pmatrix}
\xi_{\rm up} \\
\xi_{\rm lp} \\
\Lambda_{\rm up} \\
\Lambda_{\rm lp} \\
\end{pmatrix}
=
U
\begin{pmatrix}
X \\
\tilde{A} \\
\Pi \\
\tilde{D} \\
\end{pmatrix}
\end{equation}

Evidently the eigenstates of Ham. \eqref{eq:Ham_USC_diagonalised} are given by the polaritonic operators
\begin{equation}
p_{\rm up/lp} = \frac{1}{\sqrt{2 \omega_{\rm up/lp}}} \Lambda_{\rm up/lp} - i \sqrt{\frac{\omega_{\rm up/lp}}{2}} \xi_{\rm up/lp}
\end{equation}

\section{The polaritonic photon}
\label{app:polaritonic_photon}
For a moment we focus only on the cavity-dresser Hamiltonian.
Using the canonical transformation diagonalization method explained in Appendix \ref{app:polariton_canonical_transform} we can rewrite the cavity-dresser plasma Hamiltonian as
\begin{equation}\label{eq:Ham_polaritonic_cQED}
\begin{split}
& H_{c-{\rm d}} = \sum_{\textbf k}\left[ \, \hbar \omega_{\rm lp, \textbf k} \, p_{\,{\rm lp}, \textbf k}^{\dag} p_{\,{\rm lp},  \textbf k}\, + \, \hbar \omega_{\rm up, \textbf k} \, p_{\,{\rm up}, \textbf k}^{\dag} p_{\,{\rm up},  \textbf k} \right]~.
\end{split}
\end{equation}
Here $p_{\,{\rm lp},  \textbf k}$ and $p_{\,{\rm up},  \textbf k}$ represent the annihilation operators of the lower and upper polaritons, and their frequencies are given by (see Sec. \ref{app:polariton_canonical_transform} of the SM )
\begin{equation}
\omega_{\rm up/lp, \textbf k}^2 = \frac{\omega^2_{ \textbf k} + \omega_{\rm d}^2 + \Omega_{\rm d}^2}{2}  \pm \frac{1}{2}\sqrt{\left( \omega_{\rm d}^2 + \Omega_{\rm d}^2 - \omega^2_{ \textbf k} \right)^2 + 4\Omega_{\rm d}^2\omega^2_{ \textbf k}}.
\end{equation}

The new polaritonic variables represent the correct degrees of freedom to describe the cavity-dresser system, and consequently all the physical quantities must be rewritten in this basis.
Since that the coupling between the cavity and the emitter is given by 
\begin{equation}
    H_{{\rm int},e-c} = -\frac{1}{\epsilon_0}\int d^3 r \textbf{D}(\textbf r) \cdot \textbf{P}_e(\textbf r),
\end{equation}
for our aims we mainly need to transform the cavity electric displacement field, which can be rewritten following Sec. \ref{app:polariton_canonical_transform} of the SM and using the transformation in Eq. \eqref{eq:canonical_relation_between_quadratures}
\begin{equation}\label{eq:displacement_polariton_basis}
\begin{split}
    &D(\textbf r) = i \sqrt{\frac{\epsilon_0 \hbar}{2SL_c}} \sum_{\textbf k} \omega_{\textbf k} \, e^{i \textbf k \cdot \textbf{r}_{\parallel}} 
    \\
    &\times \left( \frac{\sin \theta_{\textbf k}}{\sqrt{\omega_{\rm up, \textbf k}}} p_{\,{\rm up}, \textbf k} + \frac{\cos \theta_{\textbf k}}{\sqrt{\omega_{\rm lp, \textbf k}}} p_{\,{\rm lp}, \textbf k} \right) + {\rm h.c.}.
\end{split}
\end{equation}

Using the expression for the emitter polarization density given in the main text, and assuming that the emitter is weakly doped, we can perform the rotating-wave approximation (RWA) in the emitter-cavity interaction in Eq. (1) of the main text.
%\eqref{eq:ham_plasma_emitter}. 
It is worth noticing that the RWA cannot be implemented 
%in Eq. \eqref{eq:ham_plasma_emitter} 
by only discarding the terms $\sim \left( a_{\textbf k}b_{\textbf k} \right), ~ \left(a^{\dag}_{-\textbf k}b^{\dag}_{-\textbf k}\right)$, but it requires to switch on the polaritonic picture, and considering the electric displacement field given by Eq. \eqref{eq:displacement_polariton_basis}. This is very similar to what happens in the open driven/dissipative description of USC, where one has to switch to the polaritonic picture in order to identify the positive/negative frequencies operators that, coupling to the external bath, form the correct jump operator of the system \cite{DeLiberato_PhysRevLett.112.016401, nori_USC_review, debernardis2023relaxation}.

For the sake of completeness is also worth to calculate the dresser polarization in the polariton basis, that will be useful in the next sections. It reads
\begin{equation}\label{eq:polarization_dresser_polariton_basis}
\begin{split}
    & P_{\rm d}(\textbf r) = \sqrt{\frac{\epsilon_0\hbar}{2SL_c}\frac{\Omega_{\rm d}^2}{\omega_{\rm d}}} \sum_{\textbf k} \, e^{i \textbf k \cdot \textbf{r}_{\parallel}}  
    \\
    &\times \left(\sqrt{\frac{\omega_{\rm up, \textbf k}}{\omega_{\rm d}}}\cos \theta_{\textbf k} p_{\,{\rm up}, \textbf k} + \sqrt{\frac{\omega_{\rm lp, \textbf k}}{\omega_{\rm d}}}\sin \theta_{\textbf k}p_{\,{\rm lp}, \textbf k}\right) + {\rm h.c.}
\end{split}
\end{equation}

\section{Hybridization angle and vacuum observables}
\label{app:hybridization_angle_vac_obs}

The interest in the hybridization angle $\theta_k$ is not only limited in understanding the cavity-dresser components of the polariton excitations, but it is also linked to the understanding on how the vacuum of quantum electrodynamics is modified by the presence of matter.
Indeed, by using again the canonical formalism in Sec. \ref{app:polariton_canonical_transform} of the SM, we can compute the expectation value of any observable over the USC polaritonic vacuum $|\rm vac \rangle $, defined by 
\begin{equation}
    p_{{\rm up/lp}, \textbf k}|\rm vac \rangle = 0.
\end{equation}

Using Eqs. \eqref{eq:displacement_polariton_basis}-\eqref{eq:polarization_dresser_polariton_basis} we can calculate the vacuum fluctuations of the cavity electric field considering that
\begin{equation}
    E_{\rm TM_0}(\textbf r) = \frac{D(\textbf r)-P_{\rm d}(\textbf r)}{\epsilon_0}.
\end{equation}
We thus have that
\begin{equation}
\begin{split}
    \langle {\rm vac}|  E_{\rm TM_0, \textbf k}^2   |{\rm vac}\rangle   &= 
    \frac{1}{\epsilon_0}\langle {\rm vac}|  D^2_{\textbf k} |{\rm vac}\rangle + \langle {\rm vac} |  P_{\rm d, \textbf k}^2| {\rm vac}\rangle
    \\
    = \mathcal{E}_{\textbf k}^2 \Bigg[ & \frac{\omega_{\textbf k}}{\omega_{\rm up, \textbf k}} \sin^2 \theta_{\textbf k} +  \frac{\omega_{\textbf k}}{\omega_{\rm lp, \textbf k}}\cos^2 \theta_{\textbf k} 
    \\
    &+ \frac{\Omega_{\rm d}^2}{\omega_{\rm d}^2}\left( \frac{\omega_{\rm lp, \textbf k}}{\omega_{\textbf k}} \sin^2\theta_{\textbf k} + \frac{\omega_{\rm up, \textbf k}}{\omega_{\textbf k}}\cos^2 \theta_{\textbf k}\right) \Bigg],
\end{split}
\end{equation}
where $\mathcal{E}_{\textbf k}^2 = (\hbar \omega_{\textbf k})/(2\epsilon_0 S L_c)$.
We immediately notice from the last line of this formula that the electric field fluctuations take a large contribution from the fluctuations of the dresser polarization density, which are given by the intrinsic fluctuations of matter.
Moreover, after a few algebraic steps, we have that
\begin{equation}
    \langle {\rm vac} |  P_{\rm d, \textbf k}^2 | {\rm vac}\rangle = \frac{\Omega_{\rm d}^2}{\omega_{\rm d} \omega_{\textbf k}}\langle {\rm vac}|  D^2_{\textbf k} |{\rm vac}\rangle,
\end{equation}
from which we arrive to
\begin{equation}
   \langle {\rm vac}|  E_{\rm TM_0, \textbf k}^2   |{\rm vac}\rangle   =  \frac{1}{\epsilon_0}\left[ 1 + \frac{\Omega_{\rm d}^2}{\omega_{\rm d} \omega_{\textbf k}} \right] \langle {\rm vac}|  D^2_{\textbf k} |{\rm vac}\rangle .
\end{equation}
It is worth noticing that - being a gauge non-invariant quantity - the physical significance of the electric displacement is sometimes considered obscure and confusing \cite{purcell1965berkeley}.
However, we can see here that in the dipole picture, the electric displacement is directly related to the TM$_0$ electric field fluctuations and it thus realizes a good proxy to explore the USC modifications of the electric field fluctuations.

Another interesting example that we report for completeness is the cavity virtual photon population
\begin{equation}
\begin{split}
    &N_{\rm ph, \textbf k} = \braket{{\rm vac} |\, a_{\textbf k}^{\dag} a_{\textbf k} \, | {\rm vac}} 
    \\
    &= 
    \frac{\sin^2 \theta_{\textbf k}}{4 \, \omega_{\textbf k}} \frac{\omega_{\rm up, \textbf k}^2 + \omega_{\textbf k}^2}{\omega_{\rm up, \textbf k} } + \frac{\cos^2 \theta_{\textbf k}}{4 \, \omega_{\textbf k}} \frac{\omega_{\rm lp, \textbf k}^2 + \omega_{\textbf k}^2}{\omega_{\rm lp, \textbf k}} - \frac{1}{2},
\end{split}
\end{equation}
and the bare-dresser virtual excitation population
\begin{equation}
\begin{split}
    &N_{\rm d, \textbf k} = \braket{{\rm vac} |\, d_{\textbf k}^{\dag} d_{\textbf k} \, | {\rm vac}} 
    \\
    &= \frac{\sin^2 \theta_{\textbf k}}{4 \, \omega_{\rm d}} \frac{\omega_{\rm lp, \textbf k}^2 + \omega_{\rm d}^2}{\omega_{\rm lp, \textbf k} } + \frac{\cos^2 \theta_{\textbf k}}{4 \, \omega_{\rm d}} \frac{\omega_{\rm up, \textbf k}^2 + \omega_{\rm d}^2}{\omega_{\rm up, \textbf k}} - \frac{1}{2}.
\end{split}
\end{equation}

For many years, these quantities were at the center of the discussions around polaritonic vacuum observables \cite{Ciuti_PhysRevB.72.115303, Deliberato_PhysRevLett.98.103602, deLiberato_virtualphotons}. 
However, their individual relevance is now considered marginal, since their physical meaning explicitly depends from the chosen representation \cite{Pilar_2020_thermodynamics_USC}.
They are important only when correlated with {\it physical} gauge invariant quantities.  An example of gauge-invariant quantities is the differential zero point frequency of the system $\Delta \omega_{\rm ZP}$ \cite{Ciuti_PhysRevB.72.115303}. This is obtained subtracting the bare total zero point frequency for vanishing light-matter coupling from the interacting one.
Taking the vacuum expectation value of the cavity-dresser Hamiltonian in Eq. (2) of the main text
%in Eq. \eqref{eq:Ham_c-d} 
we have that
\begin{equation}
\begin{split}
    \Delta \omega_{\rm ZP} &= \frac{\omega_{\rm up, \textbf k} + \omega_{\rm lp, \textbf k}}{2} - \frac{\omega_{\textbf k }+\omega_{\rm d}}{2} 
    \\
    &= \omega_{\textbf k}N_{\rm ph, \textbf k} + \omega_{\rm d} N_{\rm d, \textbf k} + \Omega_{\rm d}N_{\rm int, \textbf k}.
\end{split}
\end{equation}
Here the last term represents the interaction energy, defined by
\begin{equation}\label{eq:N_int}
\begin{split}
    &N_{\rm int, \textbf k} = 
    \quad \frac{i}{2} \sqrt{\frac{\omega_{\textbf{k}}}{\omega_{\text{d}}}} \langle {\rm vac} | \left(a_{\textbf{k}}^{\dag} -  a_{-\textbf{k}}^{\vphantom{\dag}}\right) \left(d_{-\textbf{k}}^{\vphantom{\dag}} + d_{\textbf{k}}^{\dag}\right) | {\rm vac} \rangle 
    \\
    &+ \langle {\rm vac} | \frac{\Omega_{\text{d}}}{4\omega_{\text{d}}}  \left(d_{\textbf{k}}^{\vphantom{\dag}} + d_{-\textbf{k}}^{\dag}\right) \left(d_{\textbf{k}}^{\vphantom{\dag}} + d_{-\textbf{k}}^{\dag}\right) | {\rm vac} \rangle .
\end{split}
\end{equation}
It is worth noticing that this term contains both the cavity-dresser interaction and the dresser self interaction, which is notoriously known as the $P^2$-term \cite{DeBernardis_cavityQED_nonperturbative_PhysRevA.97.043820, Todorov_2012_PhysRevB.85.045304, vukics_PhysRevA.94.033815, bamba_PhysRevA.90.063825, rubio_P2_ACSPh_Relevance_of_the_Quadratic_Diamagnetic}, responsible of the so-called polariton gap.

Interestingly, for the resonant wavevector $\textbf k_{\rm res}$, such that $\omega_{\textbf k_{\rm res}} = \omega_{\rm d}$, the interaction energy exactly vanishes
\begin{equation}
    N_{\rm int, \textbf k_{\rm res}} = 0,
\end{equation}
because the positive dresser self interaction ($P^2$-term) exactly compensates the negative cavity-dresser contribution in Eq. \eqref{eq:N_int}.
As a consequence the differential zero point frequency is completely determined by the cavity and dresser virtual excitation, taking the simple expression
\begin{equation}
    N_{\rm ph, \textbf k_{\rm res}} = N_{\rm d, \textbf k_{\rm res}} = \frac{1}{2}\left(\sqrt{1+\frac{\Omega_{\rm d}^2}{4\omega_{\rm d}^2}} - 1 \right).
\end{equation}
In this case the virtual photon number $N_{\rm ph, \textbf k_{\rm res}}$ represents the electromagnetic energy that can be released by an instantaneous suppression of the cavity-dresser coupling \cite{Ciuti_PhysRevB.72.115303}.

\section{Classical theory of transmission spectra}
\label{app:classical_trans_spectra}

Here we derive the linear response theory following from our cavity-dresser-emitter system.
We start by considering the total Hamiltonian given in Eq. \eqref{eq:ham_total_bosonic_rep} and rewriting it using the quadrature canonical representation for the cavity, the dresser and the emitter degrees of freedom as given in Sec. \ref{app:polariton_canonical_transform} of the SM
\begin{equation}
    H = \frac{1}{2} v^T M_{\rm tot} v,
\end{equation}
where $v^T = (X_{\rm d}(k), X_e(k), \tilde{A}_k, \Pi_{\rm d}(k), \Pi_e(k), \tilde{D}_k)$ is the array containing the canonical variables obtained generalising the definition in Eq. \eqref{eq:def_canonical_quadratures} to the emitter, while the Hamiltonian matrix is given by
\begin{equation}
M_{\rm tot} =
\begin{pmatrix}
\omega_{\rm d}^2 + \Omega_{\rm d}^2 & 0 &\Omega_{\rm d} \omega_k & 0 & 0 & 0 \\
0 & \omega_{e}^2 + \Omega_{e}^2 &\Omega_{e} \omega_k & 0 & 0 & 0 \\
 \Omega_{\rm d} \omega_k & \Omega_{e} \omega_k & \omega_{k}^2& 0 & 0 & 0 \\
 0 & 0 & 0 & 1 & 0 & 0 \\
 0 & 0 & 0 & 0 & 1 & 0 \\  
 0 & 0 & 0 & 0 & 0 & 1 \\
\end{pmatrix}
\end{equation}

From here we can write down the equation of motion of the system (using the Hamilton equations), which match the standard dielectric description from classical electromagnetism.
In the frequency domain, the equation of motion read
\begin{equation}
\mathcal{M}_k(\omega) \cdot 
    \begin{pmatrix}
        A_k\\
        X_{\rm d}(k)\\
        X_e(k)\\
    \end{pmatrix}
    = 0
\end{equation}
where the dynamical matrix is given by
\begin{equation}
\begin{split}
    &\mathcal{M}_k(\omega) =
    \\
&    
\begin{pmatrix}
\omega_k^2 - (\omega+i\gamma/2)^2 & i\omega\Omega_{\rm d} & i\omega\Omega_{e}\\
-i\omega \Omega_{\rm d} & \omega_{\rm d}^2 - (\omega+i\kappa_{\rm d}/2)^2 & -\Omega_{\rm d}\Omega_e\\
-i\omega \Omega_{e} & -\Omega_{\rm d}\Omega_e& \omega_e^2 - (\omega+i\kappa_e/2)^2\\
\end{pmatrix}
\end{split}
\end{equation}
Notice that here we introduce the cavity, dresser and emitter losses $\gamma, \kappa_{\rm d}, \kappa_e$ in a phenomenological way, just inserting a viscous damping in the equations.
The cavity transmission reads \cite{Ciuti_inout_PhysRevA.74.033811, Deliberato_PhysRevLett.98.103602}
\begin{equation}\label{eq:transmission_spectrum}
T_c(k, \omega) = \gamma \omega_k \left[\mathcal{M}^{-1}_k \right]_{00}.
\end{equation}

\section{Details on the hybridization angle measurement protocol}
\label{app:detailes_protocol}

Here we give a detailed description of the protocol to measure the cavity-dresser hybridization angle from the emitter-cavity-dresser spectrum.

\begin{figure}
    \centering
    \includegraphics[width=\columnwidth]{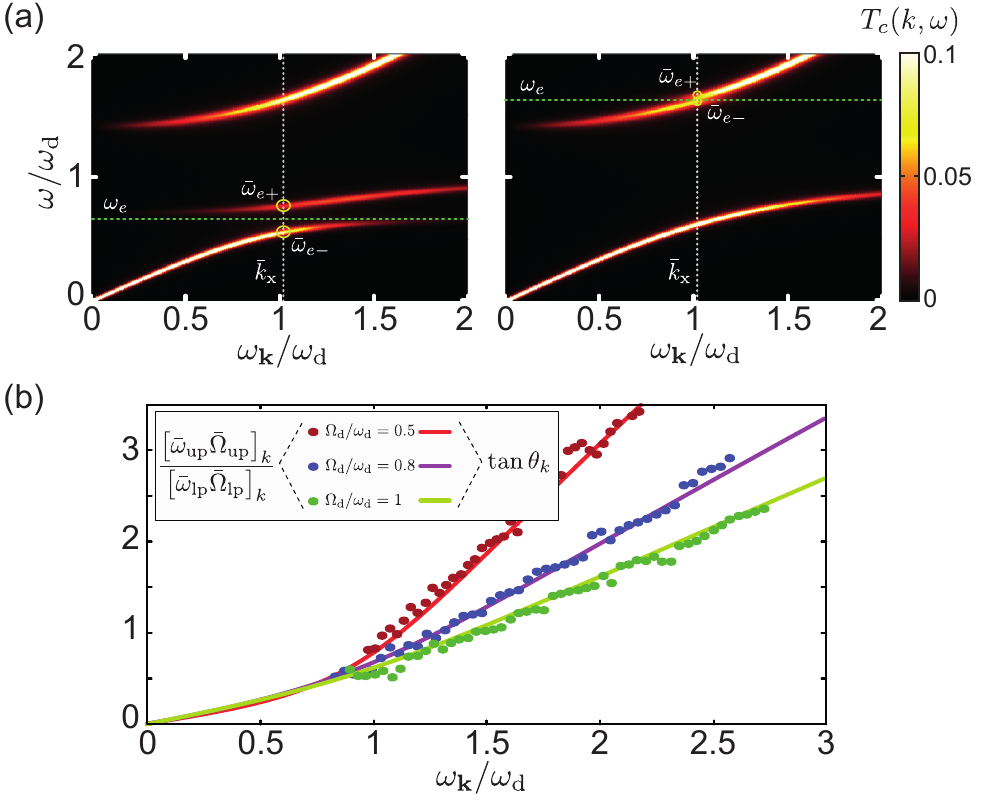}
    \caption{(a) Transmission spectrum (color scale) as a function of the incident wavevector $k$ and frequency $\omega$ calculated following Sec. \ref{app:classical_trans_spectra}. The white dashed line marks the minimal anticrossing wavevector $\bar k_{\rm x}$, the green dashed line marks the emitter resonance frequency $\omega_e$. The two yellow circle marks the two peaks where the polariton hybridizes with the emitter from which one can measure $\bar \omega_{e\pm}$.
    (b) Reconstruction of the mixing angle tangent $\tan \theta_k$ using Eq. \eqref{eq:formula_reconstruct_mixing_angle} (dots) and analytical prediction using Eq. \eqref{eq:sin_cos} from Sec. \ref{app:polariton_canonical_transform} (solid lines). The emitter frequency is swept through the two polaritons branches and every time we recorded the splitted frequencies at the minimal anticrossing in each polariton branch.  
    Parameters: (all) $\Omega_e/\omega_{\rm d} = 0.2$, $\gamma = 0.01\omega_{\rm d}$, $\kappa_{\rm d} = \kappa_{e} = 0.05\omega_{\rm d}$.
    In (a)$\Omega_{\rm d}/\omega_{\rm d}= 1$, $\omega_e/\omega_{\rm d} = 0.7 {\rm (left-panel)}-1.6 {\rm (right-panel)}$}
    \label{fig:5}
\end{figure}

We call $\bar \omega_{e+}, \bar \omega_{e-}$ the frequencies measured from the cavity transmission at the minimal anticrossing between the emitter and the cavity-dresser polaritons, while $\bar k_{\rm x}$ is the wavevector realizing the minimal anticrossing (we will keep using the bar $\bar{\cdot}$ only for quantities which are directly measured from an eventual experiment, and distinguish them from quantities derived from the theory).
We can measure $\bar k_{\rm x}, \bar \omega_{e+}, \bar \omega_{e-}$ directly from a transmission (reflection) experiment by only detecting the two peaks around the emitter frequency, as it is simulated in Fig. \ref{fig:5}(a). 
Since $\bar \omega_{e\pm} \simeq \omega_{\rm up/lp} \pm \Omega_{\rm up/lp}$, the upper/lower polariton frequency resonant with the emitter is given by
\begin{equation}
    \bar \omega_{\rm up/lp} = \frac{\bar \omega_{e+} + \bar \omega_{e-}}{2}.
\end{equation}
(Notice that the distinction between upper and lower polariton is also experimentally well defined since the two polaritons are separated by a gap, making them clearly distinguishable).
The measured emitter-polariton Rabi splitting is then given by
\begin{equation}
    \bar \Omega_{\rm up/lp} = \frac{\bar \omega_{e+} - \bar \omega_{e-}}{2}.
\end{equation}

Following Eq. (5) of the main text,
%\eqref{eq:Rabi_pol}, 
we have that the cavity-emitter anticrossing becomes a probe of the hybridization angle $\theta_{k}$, through the relation
\begin{equation}\label{eq:formula_reconstruct_mixing_angle}
    %\frac{\bar \omega_{\rm up}}{\bar \omega_{\rm lp}}\frac{\bar \Omega_{\rm up}}{\bar \Omega_{\rm lp}} = \tan \theta_{\bar k_{\rm x}}.
    \frac{\left[ \bar \omega_{\rm up}\bar \Omega_{\rm up}\right]_{k}}{\left[ \bar \omega_{\rm lp}\bar \Omega_{\rm lp}\right]_{k}} = \tan \theta_{k},
\end{equation}
where $[\cdot ]_{k}$ indicates that the data inside the square brackets are measured from the minimal anticrossing happening at the wavevector $\bar k_{\rm x} = k$.
Repeating this protocol many times while sweeping $\omega_{e}$ and collecting all the data for each emitter resonance, we can reproduce the tangent of the mixing angle using Eq. \eqref{eq:formula_reconstruct_mixing_angle}, effectively realizing a full tomography of the USC cavity-dresser polaritons.

This USC tomographic approach is naturally limited by the quality factors of the dresser, emitter and cavity $Q_{{\rm d}/e}=\omega_{{\rm d}/e} / \kappa_{{\rm d}/e}$, $Q_k = \omega_k / \gamma$, for which if the coupling between the emitter and the upper polariton at low wavevector is too small to be resolved,  $\bar \Omega_{\rm up/lp}\ll \gamma,\kappa_{{\rm d}/e}$, it is impossible to correctly identify the peaks in the transmission. 

In Fig. \ref{fig:5} we show an example of this reconstruction mechanism.
Even if here we show a specific example regarding ISB transition in a TM$_0$ cavity with linear dispersion, it is important to highlight that our formalism and the resulting tomography protocol are independent of the specific cavity QED implementation.
Not only does our description apply to any different cavity dispersions, by only inserting the specific $\omega_{\textbf k}$ in all the equations, but it applies also to any device that couples to the cavity electric field through a dipole transition.
A more detailed discussion about the generality of our analysis will be contained in Ref. \cite{DDB_GMA_coming_soon}.

In the small wavevector region of Fig. \ref{fig:5} the detection is limited by the vanishing coupling to the upper polariton, as described in Fig. 2(a)-right panel of the main text. 
%\ref{fig:2}(right panel).
In such circumstances our algorithm to reconstruct the Rabi splitting gives artificially larger data, predicting a wrong hybridization angle.
Also in the other regions our data in Fig. \ref{fig:5} are affected by numerical noise from the error committed in the peak detection due to the broadening of the transmission peaks given by the linewidths $\gamma, \kappa_{\rm d}, \kappa_e$. 
Despite that we could have completely avoided this noise by increasing our simulation's numerical accuracy, we decided to keep it in order to simulate how realistic data could be treated in an experiment and to show that our protocol works even in the presence of noisy data.

\section{Dissipations in the ultrastrong coupling polariton system}
\label{app:losses_dissipations}
Here we brefly review the theory of dissipations for the whole polaritonic system, based on the description already given in Refs. \cite{deliberato_comment_to_bamba_rates_PhysRevA.89.017801, DeLiberato_PhysRevLett.112.016401}.

Each part of the system is linearly coupled to its own independent external bath, that causes dissipations.
This interaction is also of electromagnetic origin, involving the cavity electric displacement field $D$, and the matter polarizations $P_{\rm d}, P_e$.
Following the standard assumptions for input-output theory \cite{gardiner_zoller_QNoise00}, the cavity, the dresser and the emitter couplings to their environment are described by the coupling Hamiltonians
\begin{equation}\label{eq:Ham_system_baths}
    \begin{split}
        &H_{c-B} = i\sum_{\vec k} \int d \omega \sqrt{\frac{\gamma}{2\pi} J_{\gamma }(\omega )} \left( c_{-\vec k, \,\omega} + c_{\vec k, \,\omega}^{\dag} \right)\left( a_{\vec k} - a_{-\vec k}^{\dag} \right),
        \\
        &H_{{\rm d}-B} = \sum_{\vec k} \int d \omega \sqrt{\frac{\kappa_{{\rm d}}}{2\pi}J_{\rm d }(\omega )} \left( c_{-\vec k, \,\omega} + c_{\vec k, \,\omega}^{\dag} \right)\left( d_{\vec k} + d_{-\vec k}^{\dag} \right),
        \\
        &H_{e-B} = \sum_{\vec k} \int d \omega \sqrt{\frac{\kappa_{e}}{2\pi} J_{e}(\omega ) }\left( c_{-\vec k, \,\omega} + c_{\vec k, \,\omega}^{\dag} \right)\left( b_{\vec k} + b_{-\vec k}^{\dag} \right).
    \end{split}
\end{equation}

In order to simplify the description we assumed an independent bath for each different wavevector $\vec k $, but with the same loss rates $\gamma, \kappa_{\rm d}, \kappa_e$ and  adimensional spectral densities $J_{\gamma}, J_{\rm d}, J_e$.
Furthermore, we need to express the cavity and dresser quadratures in terms of the polaritonic degrees of freedom, using the canonical representation in Appendix \ref{app:polariton_canonical_transform}.
This step is necessary in order to isolate the positive and negative frequencies component of each coupling operator to then employ the rotating-wave approximation (RWA) in Eq. \eqref{eq:Ham_system_baths} and proceed with the standard derivation.
This type of treatment is very common in the open-dissipative descriptions of all types of USC systems \cite{nori_USC_review, Solano_USC_RevModPhys.91.025005, debernardis2023relaxation}.
In this way we have
\begin{equation}
    \begin{split}
        i\left( a_{\vec k} - a_{-\vec k}^{\dag}\right) = & i\Bigg( \sin \theta_{\vec k} \sqrt{\frac{\omega_{\vec k}}{\omega_{\rm up}(\vec k)}}p_{\rm up, \, \vec k} + 
        \\
        &+\cos \theta_{\vec k} \sqrt{\frac{\omega_{\vec k}}{\omega_{\rm lp}(\vec k)}}p_{\rm lp, \, \vec k} \Bigg) + {\rm h.c.}
    \end{split}
\end{equation}
\begin{equation}
    \begin{split}
        d_{\vec k}+d_{-\vec k}^{\dag} = & i\Bigg( \cos \theta_{\vec k} \sqrt{\frac{\omega_{\rm d}}{\omega_{\rm up}(\vec k)}}p_{\rm up, \, \vec k} + 
        \\
        & - \sin \theta_{\vec k} \sqrt{\frac{\omega_{\rm d}}{\omega_{\rm lp}(\vec k)}}p_{\rm lp, \, \vec k} \Bigg) + {\rm h.c.} 
    \end{split}
\end{equation}

We can then define the polaritonic dressed loss rates as
\begin{equation}
    \begin{split}
        &\gamma_{\rm up}(\vec k ) = \gamma \sin^2 \theta_{\vec k} J_{\gamma}( \omega_{\rm up}(\vec k) ) \, \frac{\omega_{\vec k}}{\omega_{\rm up}(\vec k)},
        \\
        &\gamma_{\rm lp}(\vec k ) = \gamma \cos^2 \theta_{\vec k} J_{\gamma}( \omega_{\rm lp}(\vec k) ) \,\frac{\omega_{\vec k}}{\omega_{\rm lp}(\vec k)},
    \end{split}
\end{equation}
\begin{equation}
    \begin{split}
        &\kappa_{\rm up}(\vec k ) = \kappa_{\rm d} \cos^2 \theta_{\vec k} J_{\rm d} (\omega_{\rm up}(\vec k) ) \, \frac{\omega_{\rm d}}{\omega_{\rm up}(\vec k)},
        \\
        &\kappa_{\rm lp}(\vec k ) = \kappa_{\rm d} \sin^2 \theta_{\vec k} J_{\rm d} (\omega_{\rm lp}(\vec k) ) \, \frac{\omega_{\rm d}}{\omega_{\rm lp}(\vec k)}.
    \end{split}
\end{equation}

All the baths are assumed to be Ohmic, resulting in a adimensional spectral function $J\sim \omega/\omega_{\rm ref}$, where $\omega_{\rm ref}=\lbrace{\omega_{\vec k}, \omega_{\rm d} \rbrace}$, is a reference frequency that depends from the origin of the losses of every component, and, for simplicity, we take it equal to the resonance frequency of each component.
In this way we arrive to 
\begin{equation}
    \begin{split}
        &\gamma_{\rm up}(\vec k ) \approx \gamma \sin^2 \theta_{\vec k},
        \\
        &\gamma_{\rm lp}(\vec k ) \approx \gamma \cos^2 \theta_{\vec k} ,
    \end{split}
\end{equation}
\begin{equation}
    \begin{split}
        &\kappa_{\rm up}(\vec k ) \approx \kappa_{\rm d} \cos^2 \theta_{\vec k},
        \\
        &\kappa_{\rm lp}(\vec k ) \approx \kappa_{\rm d} \sin^2 \theta_{\vec k}.
    \end{split}
\end{equation}
One can easily verify that these expressions match the descriptions given in \cite{Bamba_Maxwell_dissipations_PhysRevA.88.013814, deliberato_comment_to_bamba_rates_PhysRevA.89.017801}.

\section{Dipolar cQED with slabs}
\label{app:dipolar_cQED_classical}

In this appendix we re-derive the whole theory in the main text starting only from Maxwell equations.
\begin{equation}\label{eq:Maxwell_inhomogeneous_div}
		\vec{\nabla}\cdot \textbf{E} = \frac{\rho}{\epsilon_0}
\end{equation}
\begin{equation}\label{eq:Maxwell_inhomogeneous_rot}
		\vec{\nabla} \times \textbf{B} = \frac{1}{c^2} \left( \frac{\textbf{J}}{\epsilon_0} + \frac{\partial}{\partial t} \textbf{E} \right)
\end{equation}

\begin{equation}\label{eq:Maxwell_structural_div}
		\vec{\nabla}\cdot \textbf{B} = 0
\end{equation}
\begin{equation}\label{eq:Maxwell_structural_rot}
		\vec{\nabla} \times \textbf{E} = - \frac{\partial}{\partial t} \textbf{B}
\end{equation}

Since we have only dipolar matter we have that
\begin{equation}
    \rho = - \vec \nabla \cdot \textbf P_{\rm tot},
\end{equation}
where $\textbf P_{\rm tot} = \sum_a \textbf P_{a}$ is the total polarization density vector of all the matter in the system, and the current is given by
\begin{equation}
    \textbf J = \partial_t \textbf P_{\rm tot}
\end{equation}

Taking the rotor of Eq. \eqref{eq:Maxwell_structural_rot} and combining with Eq. \eqref{eq:Maxwell_inhomogeneous_rot} we have that
\begin{equation}
    - \nabla^2 \textbf E + \vec \nabla (\vec \nabla\cdot \textbf E) = - \frac{1}{c^2}\partial_t^2 \textbf E - \frac{1}{c^2\epsilon_0} \partial_t^2 \textbf P_{\rm tot}.
\end{equation}
To this point, we cannot solve this equation, since it is still coupled with the Gauss law
\begin{equation}
    \vec{\nabla}\cdot \textbf{E} = -\frac{\vec \nabla \cdot \textbf P_{\rm tot}}{\epsilon_0}.
\end{equation}

We then make an arbitrary split in the electric field, defining a longitudinal and transverse part
\begin{equation}
    \textbf E = \textbf E^{\parallel} + \textbf E^{\perp},
\end{equation}
where
\begin{equation}
    \textbf E^{\parallel} = \vec \nabla( G\star \vec \nabla \cdot \textbf P_{\rm tot} )/\epsilon_0
\end{equation}
and
\begin{equation}
    \textbf E^{\perp} = \partial_t \textbf A.
\end{equation}
Here $G$ is the Green's function of the Poisson equation $-\nabla^2G(\textbf r, \textbf r') = \delta(\textbf r - \textbf r')$ with metallic boundary conditions on the plates \iac{and zero electric potential difference between them}, and $\star$ denotes the convolution operator.
Evidently $\vec \nabla \times \textbf E^{\parallel} = 0$ and the vector is longitudinal by definition, even in the standard sense \cite{Cohen-Tannoudji:113864}.
We then take $\textbf A$ as a transverse vector, having the property $\vec \nabla \cdot \textbf A = 0$.
We highlight that these definitions are general and true in a cavity setup or in any other confined geometry.

Using these definitions for the electric field we arrive at rewriting the Maxwell equations in only one equation
\begin{equation}
    - \nabla^2 \textbf E^{\perp} + \frac{1}{c^2}\partial_t^2 \textbf E^{\perp} =  - \frac{1}{c^2\epsilon_0} \partial_t^2 \left[ \textbf P_{\rm tot} + \vec \nabla( G\star \vec \nabla \cdot \textbf P_{\rm tot} )  \right].
\end{equation}
The part in square brackets on the left-hand side is the transverse projected polarization density
\begin{equation}
    \textbf P_{\rm tot}^{\perp} = \textbf P_{\rm tot} + \vec \nabla( G\star \vec \nabla \cdot \textbf P_{\rm tot} ), 
\end{equation}
with the property
\begin{equation}
    \vec \nabla \cdot \textbf P_{\rm tot}^{\perp} = 0.
\end{equation}
While the longitudinal projection is
\begin{equation}
    \textbf P_{\rm tot}^{\parallel} = - \vec \nabla( G\star \vec \nabla \cdot \textbf P_{\rm tot} )
\end{equation}

In order to understand the properties of the resulting electric field, we need to specify the dynamics of our matter system. Following the standard literature \cite{jackson_classical_1999, Hopfield_PhysRev.112.1555} we consider an equation of motion for the polarization density of each constituent of the system
\begin{equation}
    \partial_t^2 \textbf P_a +  L_a(\textbf P_a ) = \Omega_a^2\epsilon_0\textbf E (\textbf r_a ).
\end{equation}
Here $L_a(\cdot )$ is a linear differential operator defining the dynamics of the polarization density of each constituent, $\Omega_a$ is its Rabi frequency (proportional to the the plasma frequency definined in Eq. \eqref{eq:plasma_freq}).

Using all the definitions of the electric field introduced before we arrive to
\begin{equation}
    \partial_t^2 \textbf P_a +  \tilde{L}_a(\textbf P_a ) = -\Omega_a^2\sum_{b\neq a} \textbf P_{b}^{\parallel} (\textbf r_a) + \Omega_a^2\epsilon_0\textbf E^{\perp}(\textbf r_a)
\end{equation}
\begin{equation}
    - \nabla^2 \textbf E^{\perp} + \frac{1}{c^2}\partial_t^2 \textbf E^{\perp} =  - \frac{1}{c^2\varepsilon_0} \partial_t^2 \sum_a \textbf P_{a}^{\perp}(\textbf r ).
\end{equation}
Here we introduced absorbed the longitudinal self polarization term into the definition of $\tilde{L}_a$, accordingly to
\begin{equation}
    \tilde{L}_a(\textbf P_a ) = L_a(\textbf P_a ) + \Omega_a^2 \textbf P_{a}^{\parallel} (\textbf r_a).
\end{equation}
This term is the classical equivalent of the depolarization shift \cite{ando_electronic_1982}.

We now specialize in a cavity system, truncating the description to the only TM$_0$ modes \cite{Todorov_2012_PhysRevB.85.045304, Todorov_2015_dipolar_QED_2D_egas_PhysRevB.91.125409, Todorov_dipolar_QED_3D_egs_PhysRevB.89.075115}.
One can directly check that the transverse projector (or delta transverse) is given by
\begin{equation}
    \left[ \delta^{\perp}_{ij}(\textbf r , \textbf r\, ')\right]_{{\rm TM}_0} = \frac{1}{L_c}\delta_{zi}\delta_{zj}\delta(\textbf r_{\parallel} - \textbf r_{\parallel}\, ' ),
\end{equation}
here $\delta_{zi}, \delta_{zj}$ are the Kronecker deltas that select the polarization direction only along the $z$-direction, which, in our convention, is the direction perpendicular to the parallel cavity plates.
As a consequence, the longitudinal delta in the TM$_0$ mode is given by
\begin{equation}
    \left[\delta^{\parallel}_{ij}(\textbf r , \textbf r\, ')\right]_{{\rm TM}_0} = \delta_{zi}\delta_{zj}\delta(\textbf r_{\parallel} - \textbf r_{\parallel}\, ' ) \delta(z-z') - \frac{1}{L_c}\delta_{zi}\delta_{zj}\delta(\textbf r_{\parallel} - \textbf r_{\parallel}\, ' ).
\end{equation}
Notice that, when $z\neq z'$
\begin{equation}
    \left[\delta^{\parallel}_{ij}(\textbf r , \textbf r\, ')\right]_{{\rm TM}_0} = - \left[ \delta^{\perp}_{ij}(\textbf r , \textbf r\, ')\right]_{{\rm TM}_0}.
\end{equation}

Here we further specialize in the case of multiple infinite slabs, localized at different $z$-positions.
Assuming harmonic dynamics, $\tilde{L}_a(\textbf P_a )=\omega_a^2 \textbf P_a$, the equations in the TM$_0$ subspace in $k$-space (in-plane) and frequency domain reduce to
\begin{equation}\label{eq:TM0_polarizations_dyn}
    (\omega_a^2 - \omega^2 ) P_{a, k} = \Omega_a^2\sum_{b\neq a}  P_{b, k} + \Omega_a^2\epsilon_0 E_{\rm TM_0, k}
\end{equation}
\begin{equation}\label{eq:TM0_electric_field_dyn}
    (c^2 k^2 - \omega^2) E_{\rm TM_0, k} =   \frac{\omega^2}{\epsilon_0} \sum_a P_{a, k}.
\end{equation}
For simplicity from here on we suppress the vectorial notation on $k$.
Notice that similarly to the Hamiltonian description in App. \ref{app:cQED_ham_derivation}, also here the depolarization shift was absorbed into the definition of $\omega_a$.

\subsection{Effective dielectric permittivity}
To solve the equations \eqref{eq:TM0_polarizations_dyn}-\eqref{eq:TM0_electric_field_dyn} we take
\begin{equation}
    P_{a, k} = \epsilon_0 E_{\rm TM_0, k} \sum_{b} \Pi_{ab} \Omega_b^2.
\end{equation}
The dipole susceptibility is given by $\Pi_{ab} = \mathcal{M}_{ab}^{-1}$, where the dynamical matrix is
\begin{equation}
    \mathcal{M} =
    \begin{pmatrix}
        \omega_1^2 - \omega^2 && -\Omega_1^2 && -\Omega_1^2  && ... \\
        -\Omega_2^2 && \omega_2^2 - \omega^2 && -\Omega_2^2  && ... \\
        -\Omega_3^2 && -\Omega_3^2 && \omega_3^2 - \omega^2  && ... \\
        ... && ... && ... && ... \\
    \end{pmatrix}
\end{equation}

From these definitions, we find the relative electric susceptibility of the ISB multi-slabs setup as
\begin{equation}
    \epsilon_r(\omega) = 1 + \sum_{a,b} \Pi_{ab} \Omega_b^2.
\end{equation}
The cavity transmission in Eq. \eqref{eq:transmission_spectrum} is equivalently given by
\begin{equation}
    T_c(k, \omega) = \frac{\gamma \omega_k}{c^2k^2 - \omega^2\epsilon_r(\omega) - i\gamma\omega + \gamma^2/4}.
\end{equation}

For example, in the specific case of two slabs (emitter $e$ and dresser ${\rm d}$) we have
\begin{equation}
\mathcal{M} =
    \begin{pmatrix}
        \omega_{\rm d}^2 - \omega^2 && -\Omega_{\rm d}^2\\
        -\Omega_e^2 && \omega_e^2 - \omega^2\\
    \end{pmatrix}    
\end{equation}
and the dipole susceptibility is given by
\begin{equation}
\begin{split}
    \Pi = & \frac{1}{(\omega_{\rm d}^2 - \omega^2)(\omega_e^2 - \omega^2)-\Omega_{\rm d}^2\Omega_e^2} \times
    \\
    &\times 
    \begin{pmatrix}
        \omega_e^2 - \omega^2 && \Omega_{\rm d}^2\\
        \Omega_e^2 && \omega_{\rm d}^2 - \omega^2\\
    \end{pmatrix}
\end{split}
\end{equation}
As a consequence, the relative permittivity is given by
\begin{equation}
    \epsilon_r(\omega) = 1 + \frac{\Omega_{\rm d}^2(\omega_e^2 - \omega^2) + \Omega_e^2(\omega_{\rm d}^2 - \omega^2) + 2\Omega_{\rm d}^2\Omega_e^2}{(\omega_{\rm d}^2 - \omega^2)(\omega_e^2 - \omega^2)-\Omega_{\rm d}^2\Omega_e^2}
\end{equation}
It is worth noticing that when we take the limit of small Rabi (plasma) frequencies for the slabs, $\Omega_{{\rm d},e} \ll \omega_{{\rm d},e}$ we recover the usual relative permittivity for a couple of independent emitters
\begin{equation}
    \epsilon_r(\omega) \approx 1 + \frac{\Omega_{\rm d}^2}{\omega_{\rm d}^2 - \omega^2} + \frac{\Omega_e^2}{\omega_e^2 - \omega^2}.
\end{equation}

\subsection{Spontaneous emission as classical electromagnetic damping}
\label{app:spont_em}
Here we draw a connection between the modified Purcell spontaneous emission described in the maintext and the electromagnetic damping arsing in the classical equations \eqref{eq:TM0_polarizations_dyn}-\eqref{eq:TM0_electric_field_dyn}.

We start by considering the equations for the dresser and the TM$_0$ electric field mode
\begin{equation}\label{eq:dyn_dresser_E}
    \begin{split}
        & \left(\omega_{\rm d}^2 - \omega^2\right) P_{\rm d} - \Omega_{\rm d}^2\epsilon_0 E_{\rm TM_0} = \Omega_{\rm d}^2 P_e ,
        \\
        & \left(c^2k^2 - \omega^2\right) E_{\rm TM_0} - \frac{\omega^2}{\epsilon_0} P_{\rm d} = \frac{\omega^2}{\epsilon_0} P_e ,
    \end{split}
\end{equation}
coupled together to the emitter by
\begin{equation}
    \left(\omega_e^2 - \omega^2\right)P_e = \Omega_e^2\left( \epsilon_0 E_{\rm TM_0} + P_{\rm d} \right).
\end{equation}
In this last equation we immediately recognise $D = \epsilon_0 E_{\rm TM_0} + P_{\rm d}$, which is the only relevant degree of freedom that couples to the emitter.
In this section, to simplify the notation, we completely suppress the index $k$ unless necessary.

By solving Eq. \eqref{eq:dyn_dresser_E} we find that
\begin{equation}
    \epsilon_0 E_{\rm TM_0}  = \chi_{e | E}(\omega ) P_e ~~~~~~~~~~~~~~~~~~~~~~~~~~~~~~ P_{\rm d}  = \chi_{e|{\rm d}}(\omega ) P_e
\end{equation}
where the two susceptibilities are defined by
\begin{equation}
\begin{split}
    & \chi_{e | E}(\omega )  = \frac{  \omega^2\left(\bar\omega_{\rm d}^2 - \omega^2\right)  }{\left(c^2k^2-\omega^2\right)\left(\omega_{\rm d}^2-\omega^2\right) - \omega^2 \Omega_{\rm d}^2} 
    \\
    &= \omega^2 \frac{ \omega_{\rm up}^2 + \omega_{\rm lp}^2 - c^2k^2 - \omega^2 }{\left(\omega_{\rm up}^2 - \omega^2\right) \left(\omega_{\rm lp}^2 - \omega^2\right)},
    \\
    & \chi_{e|{\rm d}} (\omega ) = \frac{ \Omega_{\rm d}^2c^2k^2 }{\left(c^2k^2-\omega^2\right)\left(\omega_{\rm d}^2-\omega^2\right) - \omega^2 \Omega_{\rm d}^2} 
    \\
    &= \frac{\Omega_{\rm d}^2c^2k^2 }{\left(\omega_{\rm up}^2 - \omega^2\right) \left(\omega_{\rm lp}^2 - \omega^2\right)},
\end{split}
\end{equation}
where $\bar \omega_{\rm d}^2 = \omega_{\rm d}^2 + \Omega_{\rm d}^2$.
The emitter dynamics can be rewritten exactly as
\begin{equation}
    \left(\omega_e^2 - \omega^2\right)P_e = \Omega_e^2 \left[ \chi_{e | E}(\omega ) + \chi_{e|{\rm d}} (\omega ) \right]P_e
\end{equation}

Now we consider that the poles of $\chi_{e | E}(\omega ),\, \chi_{e|{\rm d}}(\omega )$ are lifted by the respective polariton linewidths $\gamma_{\rm up,lp}$. This is implemented by replacing $\omega^2\longmapsto \omega^2 + i\gamma_{\rm up,lp}\omega$ in the denominator of the two susceptibilities.
In this way we can specialize on the \emph{Purcell regime} where $\gamma_{\rm up,lp} \gg \Omega_e$. 
Joining this condition with the assumption that the emitter is almost resonant with one of the polaritons, for instance $\omega_e \simeq \omega_{\rm up}$, we can expand the emitter dynamics and the susceptibilities around this pole, obtaining
\begin{equation}
\begin{split}
    & \chi_{e | E}(\omega ) \approx \frac{i}{\gamma_{\rm up}} \omega_{\rm up} \frac{ c^2k^2 - \omega_{\rm lp}^2}{\omega_{\rm up}^2 - \omega_{\rm lp}^2} 
    \\
    & \chi_{e|{\rm d}} (\omega ) \approx \frac{i}{\gamma_{\rm up}} \frac{\Omega_{\rm d}^2c^2k^2}{\omega_{\rm up}}\frac{1}{\omega_{\rm up}^2 - \omega_{\rm lp}^2},
\end{split}
\end{equation}
from which
\begin{equation}
\begin{split}
    &\left(\omega_e^2 - \omega^2\right)P_e^+ \approx 2\omega_e (\omega_e - \omega) P_e^+ 
    \\
    &= \Omega_e^2 \left[ \chi_{e | E}(\omega ) + \chi_{e|{\rm d}} (\omega ) \right]P_e^+ \approx i \frac{\Omega_e^2}{\gamma_{\rm up}} \frac{\omega_{k}^2}{\omega_{\rm up}} \frac{\omega_{\rm lp}^2 - \omega_k^2}{\omega_{\rm lp}^2 - \omega_{\rm up}^2} P_e^+ ,
\end{split}
\end{equation}
where we adopt the notation $P_e^+$ to indicate that this equation describes only the dynamics of $P_e$ around the pole $\omega\simeq +\omega_e \simeq +\omega_{\rm up}$.
Rearranging the terms and taking the inverse Fourier transform, we finally finds
\begin{equation}
    i\partial_t P_e^+ = \omega_e P_e^+ - i \frac{\Omega_{\rm up}^2}{2\gamma_{\rm up}}P_e^+
\end{equation}
One can repeat the reasoning for all the poles $\omega\simeq - \omega_{\rm up}$, $\omega\simeq \pm \omega_{\rm lp}$, obtaining the same type of result.

Reinterpreting $P_e^+$ as $b_{\textbf{k}}$ we obtain the equation shown in the maintext.

\bibliographystyle{mybibstyle}
 
\bibliography{references}

\end{document}